\documentclass[%
preprint,
 amsmath,amssymb,
 aps,
]{revtex4-2}

\usepackage{graphicx}% Include figure files
\usepackage{dcolumn}% Align table columns on decimal point
\usepackage{bm}% bold math
\usepackage{braket}
\usepackage{here}
\usepackage{color}

\newcommand{\yo}[1]{{\color{black} #1}}
\begin{document}

\preprint{APS/123-QED}

\title{Irreversible energy extraction from negative temperature two-dimensional turbulence}% Force line breaks with \\

\author{Yohei Onuki}
\altaffiliation[Also at ]{Laboratoire de Physique, \'Ecole Normale Sup\'erieure de Lyon, 46 All\'ee d'Italie, F-69342 Lyon, France.}%Lines break automatically or can be forced with \\
\email{onuki@riam.kyushu-u.ac.jp}
\affiliation{%
Research Institute for Applied Mechanics, Kyushu University,\\
6-1 Kasuga-koen, Kasuga, Fukuoka, Japan
}%

\date{\today}% It is always \today, today,

\begin{abstract}
The formation and transition of patterns of two-dimensional turbulent flows observed in various geophysical systems are commonly explained in terms of statistical mechanics. Different from ordinary systems, for a two-dimensional flow, the absolute temperature defined for a statistical equilibrium can take negative values. In a state of negative temperature, the second law of thermodynamics predicts that energy in microscopic fluctuations is irreversibly converted to a macroscopic form. This study explores the possibility of this one-way energy conversion in a two-dimensional flow \yo{using a basic conceptual model}.
We consider an inviscid incompressible fluid contained in a bounded domain, the shape of which is distorted by an externally imposed force. Unlike the usual fixed boundary cases, the flow energy within the domain is exchanged with the external system via pressure work through the moving lateral boundary. Concurrently, the flow field remains constrained by vorticity conservation. Beginning from a state of Kraichnan's grand-canonical ensemble, when the domain shape is distorted from one shape to another in a finite time, the Jarzynski equality is established. This equality states that, on average, the direction of a net energy flow through the boundary during a cycle of domain distortion changes with the sign of the initial temperature of the system. Numerical experiments are carried out to verify this theoretical argument and to investigate the parameter dependence of the energy exchange rate.
\end{abstract}

\maketitle

%\tableofcontents

\section{\label{sec:Introduction}Introduction}

Two-dimensional turbulence is the simplest idealization of geophysical and astrophysical flows. When the Reynolds number is sufficiently large that viscosity is negligible, because of the vorticity conservation law, a two-dimensional flow possesses an infinite number of invariants. In particular, conservation of the second moment of vorticity (i.e., the enstrophy) has a unique role in partitioning energy across length scales. Through nonlinear interactions among turbulent eddies, the enstrophy is transferred to a smaller scale; the energy is then transferred back to a larger scale. As a consequence of this dual cascade, energy is accumulated in the largest-scale mode, resulting in the spontaneous emergence of coherent flow structures such as jets and vortices. This striking feature distinguishes two-dimensional turbulence from a three-dimensional counterpart; it has attracted considerable attention from physicists, mathematicians, and geophysical scientists. Since the middle of the 20th century, many attempts have been made to construct a theoretical basis for structure formation in two-dimensional turbulence (for representative reviews, see \citep{kraichnan_two-dimensional_1980, tabeling_two-dimensional_2002, boffetta_two-dimensional_2011}). A major direction involves the use of equilibrium statistical mechanics \citep{majda2006nonlinear, bouchet_statistical_2012, campa2014physics}.

In equilibrium statistical mechanics, the macroscopic nature of a dynamical system consisting of innumerable interacting elements is predicted. Despite the tremendous success of statistical mechanics in modern condensed matter physics, its applicability to turbulent flows is limited. It is because a microcanonical ensemble constrained by energy conservation leads to ultraviolet divergence in a wavenumber spectrum. Statistical equilibria of a three-dimensional fluid make sense only when we set an upper bound in wavenumber below which energy is partitioned \cite{cichowlas2005effective,gorce2022statistical}. If a fluid motion is constrained on a two-dimensional surface, on the other hand, enstrophy conservation restricts energy transfer in wavenumber space\textemdash energy accumulation at the largest spatial scales greatly suppresses microscopic fluctuations differently from three-dimensional cases \cite{bouchet_statistical_2012}. Consequently, equilibrium statistical mechanics readily predicts the gross nature of a continuous flow, such as the macroscopic spatial structure specified by coarse-grained streamlines. Indeed, various flow patterns observed in Earth and planetary sciences (e.g., oceanic rings and jets, the atmospheric polar vortex, and Jupiter's Great Red Spot) have been successfully explained based on equilibrium statistical mechanics \citep{michel_statistical_1994, bouchet_emergence_2002, weichman_equilibrium_2006, venaille_oceanic_2011, venaille_bottom-trapped_2012, yasuda_new_2017}.

In contrast to the typical cases in condensed matter physics, the temperature of an equilibrium state defined as the derivative of entropy with respect to energy, under the constraints of other macroscopic parameters, can be negative in a two-dimensional flow system. Using a point vortices model, \citet{onsager_statistical_1949} provided a report of the negative temperature state in a fluid. It allowed him to explain the spontaneous aggregation of same-sign vortices. In later years, many other models have been proposed to describe statistical equilibria for a broader range of fluid systems \citep{kraichnan_statistical_1975, kraichnan_two-dimensional_1980, salmon_equilibrium_1976, salmon_lectures_1998, miller_statistical_1990, robert_maximum-entropy_1991, robert_statistical_1991}, and they have commonly shown the existence of negative temperature states. However, to the author's knowledge, there remains a lack of clarity concerning specific physical properties that differ between negative and positive temperature states of fluid systems. Indeed, in the equilibrium statistical mechanics framework, the sign of temperature is not essential. When the consideration is extended into non-equilibrium cases, negative temperature reveals its peculiar characteristics; precisely, the direction of the energy flow predicted from the second law of thermodynamics is reversed.

\begin{figure}
\centering
\includegraphics[height=5cm]{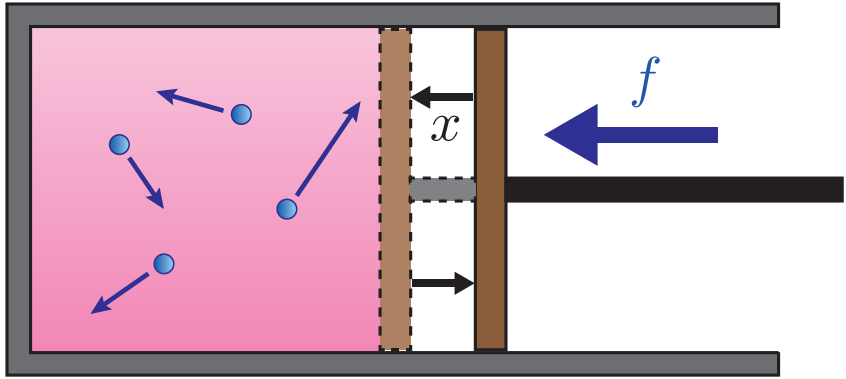}
\caption{Example of a typical thermodynamics model. A cylinder made of insulating material contains ideal gas and is connected to an external system through a piston. The piston can be moved to compress or expand the gas inside the cylinder. The piston position measured from its initial position is given by $x$, while the external force that sustains the piston is given by $f$. Then, the work performed on the gas is expressed as $W = \int f dx$. According to the first law of thermodynamics, or the law of energy conservation, $W$ should coincide with an increase in the internal energy of the gas. When the piston is moved by some distance and pulled back to its initial position, the second law of thermodynamics claims $W \geq 0$; thus, the work is irreversibly converted into thermal energy.}\label{fig:piston}
\end{figure}

In most problems, the monotonic increase of entropy claims that thermal energy contained in random and microscopic motions (i.e., heat energy) cannot be extracted macroscopically into work in any kind of machinery without changing other conditions throughout the system. Let us take a specific example; we consider ideal gas contained in a cylinder made of insulating material and connected to an external system through a movable piston (FIG.~\ref{fig:piston}). The gas is assumed to be in equilibrium at the initial time. Then, the piston is moved inward by some distance to compress the gas and pulled back to the initial position. If this experiment is performed sufficiently slowly that the gas is always in quasistatic equilibrium, the procedure is reversible; the energy of the gas in the final state is the same as that in the initial state. In contrast, if the piston is moved at a finite rate, the gas remains in non-equilibrium throughout the process resulting in increase of entropy. Consequently, energy is inevitably greater in the final state than in the initial state\textemdash the total work performed by the piston is irreversibly converted into thermal energy. This one-way energy conversion from a macroscopic form to heat is, however, dependent on the assumption that absolute temperature is positive. If the absolute temperature is negative, increase in entropy leads to the decrease in energy of the gas. Consequently, the piston extracts energy in a macroscopic form outside the system.

The \yo{dependence} of energy flow directions on a temperature sign is systematically explained from a celebrated expression derived by \citet{jarzynski_nonequilibrium_1997}. According to his formulation, work $W$ performed on a system initially attached to a thermal bath with temperature $T$ is related to the difference in the Helmholtz free energy $\Delta F$ via
\begin{equation}
\overline{\exp(- \beta W)} = \exp(- \beta \Delta F) ,
\label{eq:Jarzynski0}
\end{equation}
where $\beta \equiv 1 / k_B T$, with $k_B$ as Boltzmann's constant; the overline represents the ensemble average over all the possible initial states. The Helmholtz free energy $F$ is generally a function of the temperature $T$ and external parameters (e.g., a piston position), which are designated as $\lambda$. Then, the difference in $F$ is represented as $\Delta F = F(T, \lambda_{fin}) - F(T, \lambda_{ini})$, where $\lambda_{ini}$ and $\lambda_{fin}$ are the initial and final values of $\lambda$, respectively. Equation (\ref{eq:Jarzynski0}) is referred to as the Jarzynski equality. In the original study by Jarzynski, this equality is derived for a situation where the system is attached to a thermal bath. However, this equation is valid even for a thermally isolated system. In that case, $T$ should be interpreted as the initial temperature of the system. By application of an identity $\overline{\exp x} \geq \exp \overline{x}$, from the Jarzynski equality, the inequality $\overline{W} / T \geq \Delta F / T$ can easily be derived. If $T$ is positive, this inequality is equivalent to the well-known formula $\overline{W} \geq \Delta F$, which restricts the work available from a heat engine. In the experiment of the piston raised above, because $\lambda_{ini} = \lambda_{fin}$ leads to $\Delta F = 0$, we obtain $\overline{W} \geq 0$; accordingly, the total work performed on the system is inevitably positive. If $T$ is negative, in contrast, the inequality is reversed to yield $\overline{W} \leq 0$. That is, the energy of the randomly fluctuating motion is irreversibly extracted outside the system in a macroscopic form.

Historically, this kind of peculiarity in a state of negative absolute temperature has been investigated in a particular area of quantum mechanics, such as nuclear spins \cite{purcell1951nuclear,ramsey1956thermodynamics,hakonen1992observation,oja1997nuclear,rapp2010equilibration,braun2013negative}. It would be intriguing to explore similar problems in classical fluid dynamics. \yo{Because we are concerned with an incompressible fluid system, the typical piston model is inappropriate. Instead, we consider a closed domain whose boundary is distorted by an externally imposed force while the area is restricted to satisfy the incompressible condition. The vorticity conservation still holds for this model while the amount of energy changes because of the pressure work performed across the moving boundary\textemdash analogous to the gas-piston system that gains or loses energy while retaining the same number of molecules.

It is noted that the present model has much common with that of Gundermann et al. \cite{gundermann2013crooks}. Their interest was to verify the Crooks fluctuation theorem that connects probabilities of energy gain and loss for a pair of forward and backward processes initiated from a common thermal equilibrium \cite{crooks1999entropy}. This study, on the other hand, considers a single process where the Jarzynski equality, the integrated version of the fluctuation theorem, is relevant. We are particularly interested in temperature dependence of the energy variations during periodic distortions of the flow geometry.}

% It would be intriguing to explore similar problems in classical fluid dynamics. The primary purpose of this study is to propose a new theoretical framework that demonstrates one-way energy extraction from a negative temperature state of two-dimensional turbulent flows. Because we are concerned with an incompressible fluid system, the typical piston model is inappropriate. Instead, we consider a closed domain whose boundary is distorted by an externally imposed force while the area is restricted to satisfy the incompressible condition. A notable feature of this thought is that the enstrophy is conserved while the amount of energy changes because of the pressure work performed across the moving boundary. This situation adequately corresponds to the gas-piston system that gains or loses energy while retaining the same number of molecules. Furthermore, as we shall see, a special choice of initial statistical equilibria makes it amenable to the application of the Jarzynski's formula.

The plan of this paper is as follows. In Section~\ref{sec:formulation}, we set up a two-dimensional inviscid fluid system with a moving boundary. The flow field is separated into two parts; a boundary-induced potential flow and a vortical flow. We then analyze the response of the vortical flow part to an externally imposed potential flow part. In Section~\ref{sec:statistical}, we apply a statistical mechanics theory to this flow system and demonstrate the derivation of the Jarzynski equality. In Section~\ref{sec:experiments}, we perform numerical experiments to verify the theoretical predictions and to discuss the energy efficiency of the system quantitatively. Discussion and conclusions are presented in Section~\ref{sec:conclusions}.

\section{Formulation}\label{sec:formulation}

\begin{figure}
\centering
\includegraphics[height=5cm]{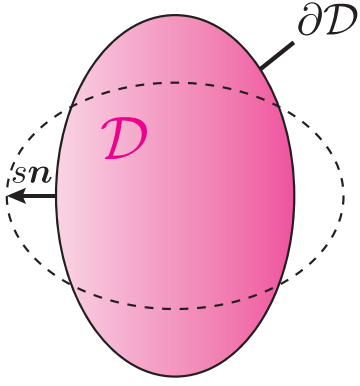}
\caption{We consider a two-dimensional fluid in a closed domain $\mathcal{D}$ with a moving boundary condition. The speed of the boundary motion, i.e, the velocity of the boundary element projected onto the unit vector normal to the boundary, $\boldsymbol{n}$, is represented as $s$.}\label{fig:ellipse}
\end{figure}

We begin the study by introducing the simplest model of an incompressible two-dimensional flow, the Euler equation:
\begin{align}
\frac{\partial \boldsymbol{u}}{\partial t} + \boldsymbol{u} \cdot \nabla \boldsymbol{u} & = - \nabla p \label{eq:Euler} \\
\nabla \cdot \boldsymbol{u} & = 0 , \label{eq:incompressible}
\end{align}
where $\boldsymbol{u}(\boldsymbol{r}, t) \equiv (u, v)$ and $p(\boldsymbol{r}, t)$ are the velocity vector and the pressure divided by density, which are functions of the spatial coordinates $\boldsymbol{r} \equiv (x, y)$ and time $t$; $\nabla \equiv (\partial_x, \partial_y)$ represents the spatial gradient operator. In the following text, we also use a subscript of ``$,x$'' or ``$,y$'' to indicate the partial derivative and an overhead dot to denote an ordinary differentiation with respect to time. The fluid is contained in a simply connected bounded domain $\mathcal{D}$, whose boundary $\partial \mathcal{D}$ moves over time. Conceptually, this boundary motion is caused by an external force that controls the pressure at each location of the boundary so that the domain shape varies in a prescribed way. We designate the expansion speed of each boundary element on $\partial \mathcal{D}$ as $s$ and impose a kinematic boundary condition for $\boldsymbol{u}$ as
\begin{equation}
\boldsymbol{u} \cdot \boldsymbol{n} = s \ \ {\rm on} \ \partial \mathcal{D} ,
\label{eq:boundary}
\end{equation}
where $\boldsymbol{n}$ is a unit vector pointing outside the domain and perpendicular to the element of $\partial \mathcal{D}$ (FIG.~\ref{fig:ellipse}). In this setting, velocity normal to the boundary is prescribed, but that tangential to the boundary is not restricted, as in the usual manner of an inviscid flow. To satisfy the incompressible condition (\ref{eq:incompressible}), the area of the domain must be conserved. The energy of the system is defined as $E = 1/2 \int_\mathcal{D} |\boldsymbol{u}|^2 d\boldsymbol{r}$. Borrowing a formula (\ref{eq:integrate_f}) in Appendix \ref{appendix A}, we understand that energy is exchanged with the external system through the boundary via the pressure work as
\begin{equation*}
\dot{E} = - \oint_{\partial \mathcal{D}} ps dl ,
\end{equation*}
where $d l$ represents an infinitesimal element of the boundary path $\partial \mathcal{D}$.

Using the incompressibility condition (\ref{eq:incompressible}), we express the velocity in the following form,
\begin{equation*}
\boldsymbol{u} = \nabla \times \psi ,
\end{equation*}
where $\nabla \times \equiv (-\partial_y, \partial_x)$ is the curl operator. Then, the vorticity is expressed as
\begin{equation*}
\omega \equiv - \partial_y u + \partial_x v  = \nabla^2 \psi .
\end{equation*}
Taking the curl of (\ref{eq:Euler}), we derive the vorticity equation,
\begin{equation}
\frac{\partial \omega}{\partial t} + \boldsymbol{u} \cdot \nabla \omega = 0 ,
\label{eq:v-eq}
\end{equation}
which shows that the vorticity $\omega$ is advected by $\boldsymbol{u}$ without changing its value along the streamline. This vorticity conservation law has the distinctive character of a two-dimensional flow. In three-dimensional flow cases, additional vortex stretching terms violate the vorticity conservation.

Now, we separate the stream function $\psi$ into two parts,
\begin{equation*}
\psi = \psi_\omega + \psi_p .
\end{equation*}
The first term is associated with vortical motion, while the second term corresponds to a potential flow induced by the moving boundary. These respective terms are defined as the solutions of the following elliptic equations:
\begin{align*}
\nabla^2 \psi_\omega & = \omega  \ \ \text{with} \ \ \psi_\omega = 0 \ \text{on} \ \ \partial \mathcal{D} \\
\nabla^2 \psi_p & = 0  \ \ \text{with} \ \ \nabla \times \psi_p \cdot \boldsymbol{n} = s  \ \ \text{on} \ \ \partial \mathcal{D} .
\end{align*}
In the same manner as the stream function, we separate the velocity into vortical and potential flow parts as $\boldsymbol{u} = \boldsymbol{u}_\omega + \boldsymbol{u}_p$, where $\boldsymbol{u}_\omega \equiv \nabla \times \psi_\omega$ and $\boldsymbol{u}_p \equiv \nabla \times \psi_p$. Because the vortical flow and the boundary-induced potential flow are orthogonal to each other (i.e., $\int_\mathcal{D} \boldsymbol{u}_\omega \cdot \boldsymbol{u}_p d\boldsymbol{r} = 0$), the energy can be separated into two parts, $E = E_\omega + E_p$, where $E_\omega \equiv (1/2) \int_\mathcal{D} |\boldsymbol{u}_\omega|^2 d\boldsymbol{r} = - (1/2) \int_\mathcal{D} \omega \psi_\omega d\boldsymbol{r}$ and $E_p \equiv (1/2) \int_\mathcal{D} |\boldsymbol{u}_p|^2 d\boldsymbol{r}$.

If $\omega$ is initially zero, $\boldsymbol{u}_p$ becomes the solution of (\ref{eq:Euler}), (\ref{eq:incompressible}), and (\ref{eq:boundary}). Clearly, $\boldsymbol{u}_p$ does not depend on any previous state of the system and is determined at each time solely from the boundary conditions. In the remaining parts of this paper, we discuss the behavior of $\boldsymbol{u}_\omega$, or identically $\omega$, affected by the externally induced flow $\boldsymbol{u}_p$. From this perspective, the energy of the system should be identified with $E_\omega$, instead of the total energy, $E_\omega + E_p$. Then, we focus on the temporal variations in $E_\omega$. Because $\boldsymbol{u}_p$ is a solution of (\ref{eq:Euler}), it can be expressed as
\begin{equation}
\frac{\partial \boldsymbol{u}_p}{\partial t} + \boldsymbol{u}_p \cdot \nabla \boldsymbol{u}_p = - \nabla p_p,
\label{eq:Euler-up}
\end{equation}
where $p_p$ is the pressure determined from the incompressibility constraint, $\nabla \cdot \boldsymbol{u}_p = 0$. Subtracting (\ref{eq:Euler-up}) from (\ref{eq:Euler}), we derive
\begin{equation}
\frac{\partial \boldsymbol{u}_\omega}{\partial t} + \boldsymbol{u} \cdot \nabla \boldsymbol{u}_\omega + \boldsymbol{u}_\omega \cdot \nabla \boldsymbol{u}_p = - \nabla (p - p_p) .
\label{eq:Euler-u-omega}
\end{equation}
Multiplying this by $\boldsymbol{u}_\omega$, using the incompressible condition $\nabla \cdot \boldsymbol{u}_\omega = 0$, and integrating the equation over $\mathcal{D}$, we obtain
\begin{equation}
\dot{E}_\omega = - \int_{\mathcal{D}} \boldsymbol{u}_\omega \cdot \left( \boldsymbol{u}_\omega \cdot \nabla \right) \boldsymbol{u}_p d \boldsymbol{r} ,
\label{eq:E-omega}
\end{equation}
where we have employed (\ref{eq:integrate_f}) and the fact that $\boldsymbol{u}_\omega \cdot \boldsymbol{n} = 0$ on the boundary. Now, (\ref{eq:E-omega}) allows a reinterpretation of the energetics of the system as follows. The boundary motion directly induces a potential flow inside the domain. The vortical flow is affected by this potential flow, gaining or losing energy through the term on the right side of (\ref{eq:E-omega}). In this manner, the energy of the system is indirectly affected by the boundary motion.

The vorticity equation (\ref{eq:v-eq}) with the formula (\ref{eq:integrate_f}) ensures that the domain-integration of a function of the vorticity, or the so-called Casimir functional, does not vary:
\begin{equation*}
\frac{d}{d t} \int_\mathcal{D} S(\omega) d\boldsymbol{r} = 0,
\end{equation*}
where $S(\omega)$ is arbitrarily chosen. In the present model, energy is exchanged with the external system through the boundary while other invariants are retained. This property contrasts the current model with previous ones in which energy conservation holds on while Casimir functionals (e.g., enstrophy) decay at small scales due to viscosity \citep{naso_statistical_2010}.

Here, we introduce the eigenfunctions $e_1(\boldsymbol{r}, t), e_2(\boldsymbol{r}, t), \ldots$ of the Laplacian operator and their corresponding eigenvalues, $\lambda_1(t), \lambda_2(t), \ldots$, as
\begin{equation*}
- \nabla^2 e_n = \lambda_n e_n   \ \ \text{with} \ \ e_n = 0 \ \ \text{on} \ \ \partial \mathcal{D} .
\end{equation*}
Here, we assume that $\{ \lambda_n \}$ are initially arranged in ascending order:
\begin{equation}
0 < \lambda_1(0) \leq \lambda_2(0) \leq \ldots ,
\label{eq:lambda-order}
\end{equation}
and $\{ e_n \}$ compose an orthonormal system:
\begin{equation*}
\int_\mathcal{D} e_n e_m d\boldsymbol{r} = \delta_{nm} .
\end{equation*}
The vorticity $\omega$ is expanded in terms of $\left\{ e_n \right\}$ as
\begin{equation}
\omega(\boldsymbol{r}, t) = \sum_n \omega_n(t) e_n(\boldsymbol{r}, t) .
\label{eq:omega-expand}
\end{equation}
Notably, the location of the boundary $\partial \mathcal{D}$ varies over time; the base functions $\left\{ e_n \right\}$ and the expansion coefficients $\left\{ \omega_n \right\}$ change accordingly. The enstrophy and energy of the system can be represented as
\begin{equation*}
\Gamma_2 \equiv \frac{1}{2} \int_{\mathcal{D}} \omega^2 d\boldsymbol{r} = \sum_n  \frac{\omega_n^2}{2}
\end{equation*}
and
\begin{equation*}
E_\omega = \sum_n \frac{\omega_n^2}{2\lambda_n} ,
\end{equation*}
respectively. The total velocity, composed of the vortical and potential flow parts, is represented as
\begin{align}
u & = \sum_n \frac{\omega_n}{\lambda_n} e_{n, y} + u_p \label{eq:u-separate} \\
v & = - \sum_n \frac{\omega_n}{\lambda_n} e_{n, x} + v_p . \label{eq:v-separate}
\end{align}
Inserting (\ref{eq:omega-expand}), (\ref{eq:u-separate}), and (\ref{eq:v-separate}) into the vorticity equation (\ref{eq:v-eq}), multiplying $e_n$ and integrating it over $\mathcal{D}$, we obtain the vorticity equation expanded onto mode space as a set of ordinary differential equations,
\begin{equation}
\dot{\omega}_n = \sum_m A_{nm} \omega_m + \sum_{m, l} B_{nml} \omega_m \omega_l,
\label{eq:v-eq-mode}
\end{equation}
where the mode-coupling coefficients are defined as
\begin{align}
A_{nm} & = - \int_\mathcal{D} e_n \left( \frac{\partial}{\partial t} + \boldsymbol{u}_p \cdot \nabla \right) e_m d\boldsymbol{r} \label{eq:coupling_A} \\
B_{nml} & = \frac{1}{2 \lambda_m} \int_\mathcal{D} (e_{m,x} e_{l,y} e_{n} - e_{m,y} e_{l,x} e_{n}) d\boldsymbol{r} + (m \leftrightarrow l) . \label{eq:coupling_B}
\end{align}
These coefficients satisfy $A_{mn} = - A_{nm}$, $B_{nml} = B_{nlm}$, $B_{nml} + B_{mln} + B_{lnm} = 0$, and $B_{nml} / \lambda_n + B_{mln} / \lambda_m + B_{lnm} / \lambda_l = 0$. Accordingly, we derive the enstrophy conservation law,
\begin{equation*}
\dot{\Gamma}_2 = 0 ,
\end{equation*}
and the energy equation,
\begin{equation}
\dot{E}_\omega = \sum_{m, n} \left( - \frac{\dot{\lambda}_n \delta_{nm}}{2 \lambda^2_n} + \frac{A_{nm}}{\lambda_n} \right) \omega_n \omega_m .
\label{eq:energy-mode-total}
\end{equation}
Again, if the domain boundary $\partial \mathcal{D}$ is fixed, $\dot{\lambda}_n$ and $A_{nm}$ are identically zero; thus, the right-hand side of (\ref{eq:energy-mode-total}) vanishes. The role of the moving boundary is to cause variations in the eigenvalues and eigenfunctions, both of which serve as sources of energy in the system. When we define the energy density for each mode as $E_n = \omega_n^2 / (2 \lambda_n)$, a detailed energy equation,
\begin{equation*}
\dot{E}_n = - \frac{\dot{\lambda}_n}{\lambda_n} E_n + \sum_m \frac{A_{nm}}{\lambda_n} \omega_n \omega_m + \sum_{m, l} \frac{B_{nml}}{\lambda_n} \omega_n \omega_m \omega_l ,
\end{equation*}
is also derived.

\subsection*{Rapid distortion theory}\label{rapid-distortion-theory}

As described above, the evolution of the vorticity distribution in the present model is governed by the external and internal processes; the external process indicates advection by the boundary-induced potential flow, and the internal process indicates nonlinear interaction among vortices. The relative importance of these two processes is quantified by introducing the concept of time scales. If the time scale of the boundary motion, represented as $\tau_b$, is much shorter than the eddy overturning time scale, $\tau_e$, the flow field is dominated by the potential flow part. Thus, we can approximate $\boldsymbol{u} \sim \boldsymbol{u}_p$. Consequently, the vorticity is almost passively advected by the boundary-induced flow. The vorticity equation (\ref{eq:v-eq-mode}) is now written as
\begin{equation}
\frac{\partial \omega}{\partial t} + \boldsymbol{u}_p \cdot \nabla \omega = 0
\label{eq:v-eq-RDT}
\end{equation}
or
\begin{equation}
\dot{\omega}_n = \sum_m A_{nm} \omega_m .
\label{eq:v-eq-RDT-mode}
\end{equation}
This approximation is sometimes referred to as rapid distortion theory (RDT) \citep{savill_recent_1987, hunt_rapid_1990}. Because (\ref{eq:v-eq-RDT}) and (\ref{eq:v-eq-RDT-mode}) are linear equations, it is more tractable than the original nonlinear equation. Furthermore, in some examples that we will provide below, the boundary-induced potential flow $\boldsymbol{u}_p$ is analytically obtained. In that case, (\ref{eq:v-eq-RDT}) and (\ref{eq:v-eq-RDT-mode}) can easily be integrated from an arbitrary initial condition to predict the linear response of the vorticity field to an imposed boundary motion.

\subsubsection{Example I: Pure straining}\label{sec:strain}

\begin{figure}
\centering
\includegraphics[height=7cm]{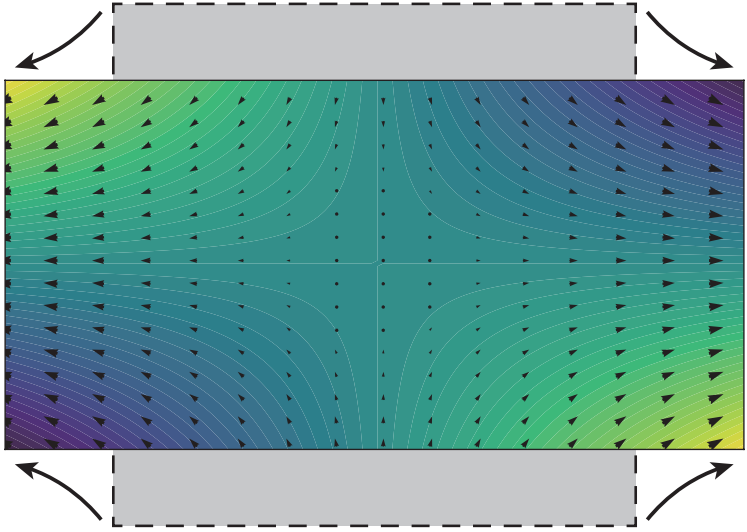}
\caption{Potential flow induced by boundary motion in a rectangular domain shrinking vertically and stretching horizontally. The color and arrows represent the stream function, $\psi_p$, and the velocity, $\boldsymbol{u}_p = \nabla \times \psi_p$, respectively.}\label{fig:strain}
\end{figure}

We consider a situation where a domain is stretched in one direction and shrunk in another (FIG.~\ref{fig:strain}). We assign coordinates $x$ and $y$ to these directions and assume they are perpendicular to each other. The strain rate is specified by a single parameter $a(t)$, which represents the aspect ratio of a small rectangular element embedded in the domain and distorted by the boundary-induced flow, $\boldsymbol{u}_p = \nabla \times \psi_p$. The stream function of the potential flow, $\psi_p$, is now written as
\begin{equation}
\psi_p(x, y) = - \frac{\dot{a}}{2a} xy + c_1x + c_2y.
\label{eq:pure-strain}
\end{equation}
Here, the second and third terms on the right-hand side represent a spatially homogeneous flow. By properly choosing the origin of the system, we can eliminate these terms without loss of generality. Hereafter, we set $c_1 = c_2 = 0$.

In this example, the linearized vorticity equation (\ref{eq:v-eq-RDT}) becomes
\begin{equation}
\partial_t \omega + \frac{\dot{a}}{2a} \left( x \partial_x \omega - y \partial_y \omega \right) = 0 .
\label{eq:v-eq-strain}
\end{equation}
When we introduce a new coordinate as $(x', y') = (x / \sqrt{a}, \sqrt{a}y)$, (\ref{eq:v-eq-strain}) reduces to $\left. \partial_t \omega \right\vert_{x', y'} = 0$. Thus, the vorticity distribution is fixed in this straining frame. The general solution of (\ref{eq:v-eq-strain}) with an arbitrary initial condition, $\omega(x,y,0) = \omega_0(x,y)$, is accordingly
\begin{equation*}
\omega(x, y, t) = \omega_0 (x / \alpha, \alpha y) ,
\end{equation*}
where $\alpha = \sqrt{a(t) / a(0)}$ is defined. Notably, the vorticity distribution depends only on its initial state and the instantaneous value of $\alpha$. Here, $\alpha$ is independent of the rate of distortion; whether we vary $a$ slowly or fast, as far as the RDT is valid and the final value of $a$ is fixed, the system reaches an identical state. This behavior is analogous to the distortion of an elastic medium.

\subsubsection*{Example II: Pure rotation}\label{example-ii-pure-rotation}

\begin{figure}
\centering
\includegraphics[height=7cm]{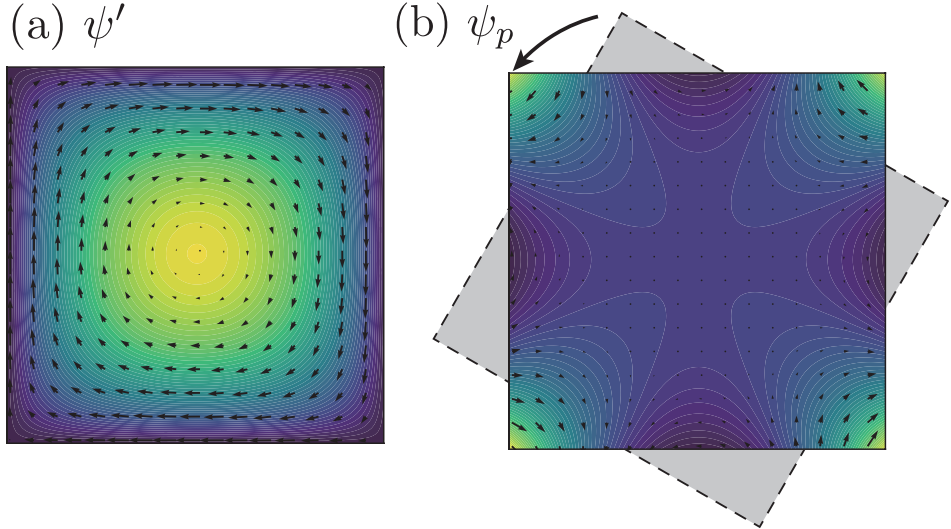}
\caption{Potential flow induced by boundary motion in a square domain rigidly rotating counterclockwise. (a) The stream function $\psi'$ and the velocity $\nabla \times \psi'$ in the rotating frame are represented by color and arrows, respectively. (b) The stream function and velocity in the fixed frame, $\psi_p$ and $\boldsymbol{u}_p = \nabla \times \psi_p$.}\label{fig:rotation}
\end{figure}

We consider a situation where a domain boundary is rotated in one direction without changing its shape (FIG.~\ref{fig:rotation}). We introduce a parameter $\theta(t)$ that represents the angle of the domain measured from the initial state. For simplicity, we regard the origin of the coordinate $(x,y) = (0,0)$ as the center of the rotation. Although the domain boundary moves rigidly, the fluid within it is not required to rotate similarly. Because the potential flow, $\psi_p$, is not associated with any vorticity, we must solve the Poisson equation to obtain a non-trivial solution. For this purpose, we consider a time-dependent coordinate change, $x' = x \cos\theta + y \sin\theta$ and $y' = - x \sin\theta + y \cos\theta$. Even in the $(x', y')$ system, the vorticity equation (\ref{eq:v-eq}) remains the same form, whereas the vorticity in the rotating frame, $\omega'$, differs from the vorticity in the fixed frame, $\omega$, as $\omega' = \omega - 2 \dot{\theta}$. To obtain the potential flow induced by the moving boundary, we assume that the vorticity is zero in the fixed frame (i.e., $\omega = 0$), which is translated to $\omega' = - 2 \dot{\theta}$ in the rotating frame. Because the domain boundary does not move in the rotating frame, the equation to be solved reduces to a simple form:
\begin{equation*}
\nabla^2 \psi' = - 2 \dot{\theta}\ \ \text{with} \ \ \psi' = 0 \ \ \text{on} \ \ \partial \mathcal{D} .
\end{equation*}
By expanding $\psi'$ onto the Laplacian eigenfunctions, we arrive at
\begin{equation*}
\psi' = 2\dot{\theta} \sum_n \frac{\gamma_n}{\lambda_n} e_n ,
\end{equation*}
where $\gamma_n = \int_{\mathcal{D}} e_n d\boldsymbol{r}$ is defined. The stream function, $\psi_p$, in the fixed frame is finally obtained as $\psi_p = \psi' + \dot{\theta} (x^2 + y^2) / 2$.

Like the pure straining case, the vorticity distribution at some instance is determined by its initial state and the rotation angle $\theta$; it is independent of the angular velocity, $\dot{\theta}$. Notably, $\theta$ covers the whole range of real numbers; i.e., $\theta + 2\pi$ should not be identified with $\theta$. The vorticity distribution will not return to the initial state even when the domain rotates over one revolution to arrive at the initial angle. This result may be visually understood from FIG.~\ref{fig:rotation}b. The potential flow is intense at the edge of the domain but very weak near the center. Therefore, vorticity is transferred much faster in the outer region than in the inner region. Although potential flow does not cause vortical motion, it causes a net differential rotation of fluid elements when averaged over time.

\section{Statistical mechanics}\label{sec:statistical}
We shall apply a theory of statistical mechanics to the flow system constructed in the previous section. At this stage, it is important to note that there are several ways to define a statistical equilibrium in an Euler equation system. The most widely accepted model for this problem is that formulated by Miller, Robert and Sommeria \cite{miller_statistical_1990, robert_maximum-entropy_1991, robert_statistical_1991}. This model, currently referred to as MRS theory, specifies coarse-grained streamlines as well as probability distributions of vorticity at a microscopic scale based on infinite number of constraints. Despite its generality, MRS theory is not best suited for the present discussion. It is because MRS theory uniquely determines the macroscopic flow structure and normally does not allow turbulent fluctuations essential to extend theory to non-equilibrium problems. The easiest way to incorporate fluctuations is provided by the Kraichnan's approach \cite{kraichnan_statistical_1975, kraichnan_two-dimensional_1980}, which we employ here. This classical model specifies the probability density functions of vorticity expanded on Laplacian eigenfunctions based only on the energy and enstrophy constraints. Limitations of the present formulation will be discussed in section \ref{sec:conclusions}.

Hereafter, we truncate the normal mode expansion (\ref{eq:omega-expand}) at a finite number, $n=N$. The state of the system is thus represented by $N$ real variables, $\{ \omega_1, \omega_2, \ldots, \omega_N \} \equiv \boldsymbol{\omega}$, which specify a point in $N$-dimensional phase space. The trajectory of this point is determined by its initial position and the time-dependent external parameters, $A_{nm}$ and $B_{nml}$, which control the governing equation, (\ref{eq:v-eq-mode}). Then, we introduce the concept of a statistical ensemble; we consider an innumerable number of points that move independently in phase space. Instead of examining the trajectory of each point, we focus on changes in the number density of points at each location fixed in phase space. When this number density is written as $P(\boldsymbol{\omega})$, its temporal variation is described by the Liouville equation:
\begin{equation}
\frac{\partial P}{\partial t} + \sum_{n=1}^N \frac{\partial P \dot{\omega}_n}{\partial \omega_n} = 0 ,
\label{eq:Liouville}
\end{equation}
where $\dot{\omega}_n$ represents the $n$th component of ``flow velocity'' of points at each location, $\boldsymbol{\omega}$, and is defined by the (truncated) vorticity equation, (\ref{eq:v-eq-mode}). Here, because $A_{nn} = 0$ and $B_{nmn} = 0$, it follows that
\begin{equation}
\frac{\partial \dot{\omega}_n}{\partial \omega_n} = 0 ,
\label{eq:DLT}
\end{equation}
and accordingly,
\begin{equation}
\sum_{n=1}^N \frac{\partial \dot{\omega}_n}{\partial \omega_n} = 0 .
\label{eq:Liouville-theorem}
\end{equation}
This equation allows (\ref{eq:Liouville}) to be rewritten as
\begin{equation}
\frac{\partial P}{\partial t} + \sum_{n=1}^N \dot{\omega}_n \frac{\partial P}{\partial \omega_n} = 0 .
\label{eq:Liouville2}
\end{equation}
Equation (\ref{eq:Liouville-theorem}) is known as Liouville's theorem, which ensures that the volume element in phase space passively advected by the vector field $\dot{\boldsymbol{\omega}}$ does not change over time. Consequently, because the number of particles contained in that element is also conserved, the number density $P$ will not vary along the trajectories. This nature enables us to utilize the well-established methods to define statistical equilibria.

In the truncated system, among the infinite number of invariant quantities in the original Euler equation, only the enstrophy,
\begin{equation}
\Gamma_2(\boldsymbol{\omega}) = \sum_{n=1}^N \frac{\omega_n^2}{2} ,
\label{eq:enstrophy-truncate}
\end{equation}
is strictly conserved. In addition, if the domain boundary is fixed, the total energy,
\begin{equation}
E_\omega(\boldsymbol{\omega}) = \sum_{n=1}^N \frac{\omega_n^2}{2 \lambda_n} ,
\label{eq:energy-truncate}
\end{equation}
is also conserved. If the probability density is a function of these invariants (i.e., written as $P(\boldsymbol{\omega}) = \mathcal{P}(E_\omega, \Gamma_2)$), it follows that $\partial P / \partial t = 0$. In this manner, based on $E_\omega$ and $\Gamma_2$, various kinds of stationary solutions of the Liouville equation (\ref{eq:Liouville2}) can be arranged. Among these arrangements, we employ the grand-canonical ensemble (GCE) that is defined by
\begin{equation}
P(\boldsymbol{\omega}) = \frac{1}{Z} e^{-N(\beta E_\omega + \alpha \Gamma_2)} ,
\label{eq:micro-measure}
\end{equation}
where $\alpha$ and $\beta$ are constants independent of $\boldsymbol{\omega}$. The coefficient $N$ on the exponential factor is absent in the usual form of GCE. This coefficient is introduced here for the purpose of making $\alpha$ and $\beta$ finite, even in the limit of $N \to \infty$, as we will describe later. The normalization constant $Z$ is the partition function defined as
\begin{equation}
Z(\alpha, \beta) \equiv \int \prod_{n=1}^N d\omega_n e^{-N(\beta E_\omega + \alpha \Gamma_2)} .
\label{eq:partition-function}
\end{equation}
By inserting (\ref{eq:enstrophy-truncate}) and (\ref{eq:energy-truncate}) into (\ref{eq:micro-measure}), $P(\boldsymbol{\omega})$ can be rewritten as a product of mutually independent Gaussian functions:
\begin{equation}
P(\boldsymbol{\omega}) = \frac{1}{Z} \prod_{n} e^{-\omega_n^2 / (2 \sigma_n^2)} ,
\label{eq:micro-measure2}
\end{equation}
with the variance for the $n$th mode defined as
\begin{equation}
\sigma^2_n = \frac{\lambda_n}{N(\beta + \alpha \lambda_n)} .
\label{eq:sigma2}
\end{equation}
Notably, $\sigma^2_n$ coincides with twice the expectation value of the enstrophy contained in the $n$th mode, or $\overline{\omega_n^2}$. In general, the eigenvalues of the Laplacian operators increase with their indices as
\begin{equation}
\lambda_n \sim O(n) .
\label{eq:lambda-scale}
\end{equation}
For $\sigma_n^2$ to be positive, even for a sufficiently large $n$, $\alpha$ must be positive or $O(1/n)$ with the negative sign. For simplicity, we assume that $\alpha$ is always positive. Then, by redefining $(\omega \alpha^{1/2}, t / \alpha^{1/2}, s \alpha^{1/2}, \beta / \alpha)$ as $(\omega, t, s, \beta)$, respectively, we can set $\alpha = 1$ without loss of generality. In addition, again for (\ref{eq:sigma2}) being positive for any $n$, $\beta > - \lambda_1$ must be held.

Now, for a given set of $\{ \lambda_n \}$ and $N$, the state of the system is specified by a single parameter $\beta$ known as the inverse temperature. Generally, $\beta$ is related to the absolute temperature, $T$, by the expression, $\beta \equiv 1/(k_B T)$, where $k_B$ is Boltzmann's constant. Therefore, the sign of $\beta$ coincides with that of temperature. In most cases, the normalization condition of $P$ allows only the positive sign of $\beta$. However, in the present case, $\beta$ ranges from negative to positive values. The existence of negative temperature states is a peculiar characteristic inherent in a two-dimensional flow system.

Because $\overline{\omega^2_n} = \lambda_n / N(\beta + \lambda_n)$ derives from (\ref{eq:sigma2}), the expectation values of energy and enstrophy are calculated from (\ref{eq:energy-truncate}) and (\ref{eq:enstrophy-truncate}) as
\begin{equation}
\overline{E}_\omega =  \frac{1}{2 N} \sum_{n=1}^N \frac{1}{\beta + \lambda_n}
\label{eq:energy-exp}
\end{equation}
and
\begin{equation}
\overline{\Gamma}_2 = \frac{1}{2 N} \sum_{n=1}^N \frac{\lambda_n}{\beta + \lambda_n} ,
\label{eq:enstrophy-exp}
\end{equation}
respectively. If we increase $N$ while fixing $\beta$, from (\ref{eq:lambda-scale}) and (\ref{eq:energy-exp}), the energy decreases as $\overline{E}_\omega \sim O(\log N / N)$ and converges to $0$ in the thermodynamic limit, $N \to \infty$. To avoid this unphysical consequence, we also vary $\beta$ in response to $N$ as
\begin{equation*}
\beta =  -\lambda_1 + \frac{1}{2 \overline{E}_1 N},
\end{equation*}
which allows (\ref{eq:energy-exp}) to be rewritten as
\begin{equation}
\overline{E}_\omega = \overline{E}_1 + \frac{1}{2 N} \sum_{n=2}^N \frac{1}{\lambda_n - \lambda_1 + 1 / (2 \overline{E}_1 N)} .
\label{eq:energy-exp2}
\end{equation}
The second term on the right side will vanish if we extend the summation to infinity, $N \to \infty$. Specifically, all the energy is concentrated in the lowest mode in the thermodynamic limit. The enstrophy distributed in each mode becomes $O(1/N)$, except for the lowest mode to which a finite value $\lambda_1 \overline{E}_1$ is assigned. The contribution from $n>1$ modes to the total enstrophy is $(1/2N) \sum_{n=2}^N (\lambda_n / (\lambda_n - \lambda_1 + (1/2\overline{E}_1 N)))$, which remains finite even in the thermodynamic limit, $N \to \infty$. Note that these results are identical to those of the microcanonical ensemble \citep{bouchet_invariant_2010}.

The partition function $Z$ defined as (\ref{eq:partition-function}) is made explicit as
\begin{equation}
Z = \left( (2\pi)^N \prod_{n=1}^N \frac{\lambda_n}{N(\beta + \lambda_n)} \right)^{1/2} .
\label{eq:partition-function2}
\end{equation}
Accordingly, the Helmholtz free energy $F(\beta)$ is calculated as
\begin{equation*}
F = -  \frac{1}{N \beta} \ln Z = F_1 (\beta; N) + F_2 (\beta; N, \lambda) ,
\end{equation*}
where
\begin{align*}
F_1 & \equiv - \frac{1}{2\beta} \ln \left( \frac{2\pi}{N} \right) \\
F_2 & \equiv \frac{1}{2 N \beta} \sum_{n=1}^N \left( \ln(\beta + \lambda_n) - \ln \lambda_n \right)
\end{align*}
are defined. Here, $F_1$ depends only on $\beta$ and $N$. In contrast, $F_2$ involves $\beta$, $N$, and the Laplacian eigenvalues, $\{ \lambda_n \}$; therefore, it depends on the domain geometry.

\subsection*{Jarzynski equality}\label{jarzynski-equality}

Here, we consider a GCE specified by the inverse temperature $\beta$ at a time $t=t_1$. Beginning from this ensemble, we move the model boundary until $t=t_2$, such that the Laplacian eigenvalues vary as $\boldsymbol{\lambda}(t) = (\lambda_1, \lambda_2, \ldots, \lambda_N)$. The probability distribution in phase space, $\boldsymbol{\omega} \equiv (\omega_1, \omega_2, \ldots, \omega_N)$, at each time $t$ is defined as $P(\boldsymbol{\omega}, t)$. The functional form of the energy of the system explicitly involves the external parameter $\boldsymbol{\lambda}$ as $E_\omega = E_\omega(\boldsymbol{\omega}; \boldsymbol{\lambda})$, while the functional form of the enstrophy depends only on $\boldsymbol{\omega}$ as $\Gamma_2(\boldsymbol{\omega})$. We assume a trajectory in phase space from $t=t_1$ to $t=t_2$ and designate the initial and final states as $\boldsymbol{\omega}(t_1) = \boldsymbol{\omega}_1$ and $\boldsymbol{\omega}(t_2) = \boldsymbol{\omega}_2$, respectively. During a process, the external parameter $\boldsymbol{\lambda}$ is switched from $\boldsymbol{\lambda}(t_1) = \boldsymbol{\lambda}_1$ to $\boldsymbol{\lambda}(t_2) = \boldsymbol{\lambda}_2$. Because the work performed on the system through the domain boundary from $t=t_1$ to $t=t_2$ is the difference in energy between these two times, it can be written as $W \equiv E_\omega(\boldsymbol{\omega}_2; \boldsymbol{\lambda}_2) - E_\omega(\boldsymbol{\omega}_1; \boldsymbol{\lambda}_1)$. In contrast to the energy that varies, the enstrophy is conserved throughout the process; $\Gamma_2(\boldsymbol{\omega}_1) = \Gamma_2(\boldsymbol{\omega}_2)$. In the context of these considerations, we evaluate the following expression:
\begin{equation*}
\overline{\exp(- N\beta W)} \equiv \int d\boldsymbol{\omega}_2 P(\boldsymbol{\omega}_2, t_2) \exp\left(- N \beta W \right).
\end{equation*}
The Liouville theorem assures conservation of the probability density along a trajectory in phase space, $P(\boldsymbol{\omega}_2, t_2) = P(\boldsymbol{\omega}_1, t_1)$, which therefore results in
\begin{align*}
P(\boldsymbol{\omega}_2, t_2) \exp (N \beta E_\omega(\boldsymbol{\omega}_1; \boldsymbol{\lambda}_1)) & = P(\boldsymbol{\omega}_1, t_1) \exp (N \beta E_\omega(\boldsymbol{\omega}_1; \boldsymbol{\lambda}_1)) \\
& = \frac{1}{Z(\beta; \boldsymbol{\lambda}_1)} \exp (-N \Gamma_2(\boldsymbol{\omega}_1)) \\
& = \frac{1}{Z(\beta; \boldsymbol{\lambda}_1)} \exp (-N \Gamma_2(\boldsymbol{\omega}_2)) .
\end{align*}
Consequently, we obtain
\begin{align}
\overline{\exp(- N\beta W)} \nonumber & = \frac{1}{Z(\beta; \boldsymbol{\lambda}_1)} \int d\boldsymbol{\omega}_2 \exp (-N (\beta E_\omega(\boldsymbol{\omega}_2; \boldsymbol{\lambda}_2) + \Gamma_2(\boldsymbol{\omega}_2))) \nonumber \\
& = \frac{Z(\beta; \boldsymbol{\lambda}_2)}{Z(\beta; \boldsymbol{\lambda}_1)} \nonumber \\
& = \exp(- N \beta \Delta F) ,
\label{eq:Jarzynski}
\end{align}
where $\Delta F \equiv F(\beta, \boldsymbol{\lambda}_2) - F(\beta, \boldsymbol{\lambda}_1)$. This result, $\overline{\exp(- N\beta W)} = \exp(- N \beta \Delta F)$, is the Jarzynski equality proposed for the first time in 1997 by \citet{jarzynski_nonequilibrium_1997}. Although it was originally derived for a canonical Hamiltonian system, the Liouville property enables establishment of this type of equation even in non-canonical systems. Importantly, (\ref{eq:Jarzynski}) is valid for a non-equilibrium process. As far as the system is initially in a (grand-)canonical ensemble state, even though the distribution function $P$ at a later time is not a stationary solution of the Liouville equation, the present formulation is exact. The Jarzinski equality is a powerful extension of equilibrium statistical mechanics to a non-equilibrium theory.

Because of the convexity of the exponential function, which assures $\overline{\exp X} \geq \exp \overline{X}$ for any random variable $X$, we can derive the following inequality:
\begin{equation}
-\beta \Delta F \geq - \beta \overline{W}.
\label{eq:inequality1}
\end{equation}
When $\beta$ is positive, which applies to most systems, (\ref{eq:inequality1}) is rewritten as $\overline{W} \geq \Delta F$. This inequality restricts the lower bound of the work exerted on the system (when $\Delta F$ is negative, the inequality restricts the upper bound of the work extracted from the system) and is recognized as a form of the second law of thermodynamic. For two-dimensional turbulence, as demonstrated in the last subsection, $\beta$ can also be negative. In that case, we obtain
\begin{equation}
\overline{W} \leq \Delta F,
\label{eq:inequality2}
\end{equation}
which asserts the upper bound of the work exerted on the system. The most striking result is derived when the shape of the domain boundary at the final time coincides with the shape of the domain boundary at the initial time. In this case, because $\boldsymbol{\lambda}_1 = \boldsymbol{\lambda}_2$ holds, $\Delta F$ becomes 0. Therefore, (\ref{eq:inequality2}) reduces to $\overline{W} \leq 0$; specifically, the expectation value of energy injected into the system becomes non-positive for any type of boundary motion. The energy of two-dimensional flows, starting from the equilibrium state of GCE with negative temperature, is on average irreversibly extracted to the outer region.

\section{Numerical analysis}\label{sec:experiments}

Theoretical considerations based on statistical mechanics have provided some clarity regarding the peculiar nature of two-dimensional flow systems. Next, to extend our understanding on a quantitative level, we compute the energy spectrum corresponding to the GCE in a specific situation and carry out numerical simulations that directly solve the Euler equation in a distorting domain. For ease of analysis, we adopt a simple model configuration of a rectangular domain periodically distorted by straining boundary motion.

Although an actual continuous fluid system inherently involves an infinite number of eigenmodes, numerical investigation forces truncation of the mode expansion at a finite number. We thus abandon the idea of the thermodynamic limit and allow both positive and negative temperature states to exist. This somewhat artificial configuration rather deserves investigation because of the following points.

In the thermodynamic limit, all energy is concentrated in the lowest mode. Consequently, the system is not associated with any velocity fluctuation. In this case, externally imposed boundary motion may cause transition of the flow field only in a coherent manner. A statistical mechanics approach is therefore unsuited for this scenario. Finite truncation of mode numbers is a better approach to discussing the statistical behavior of a turbulently fluctuating system.

\subsection{Model spectrum in a rectangular domain}\label{model-spectrum-in-a-rectangular-domain}

\begin{figure}
\centering
\includegraphics[height=8cm]{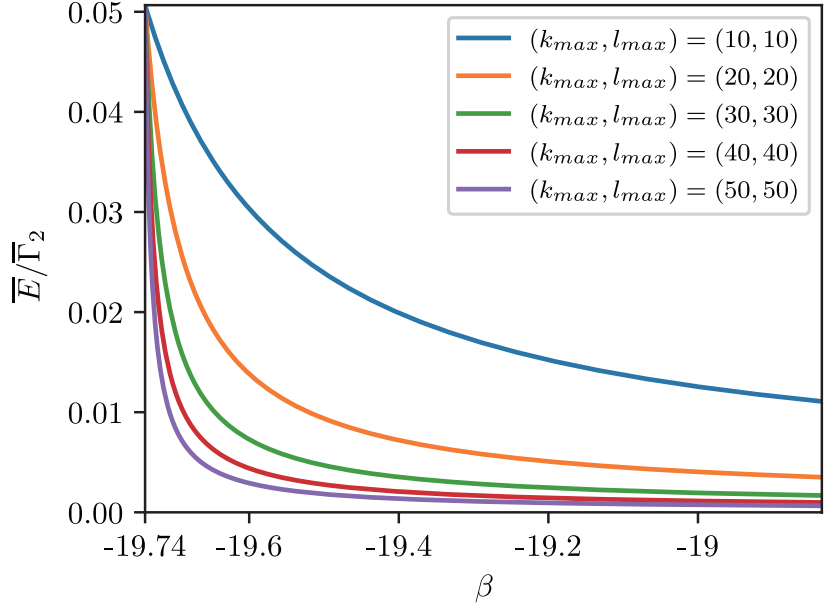}
\caption{The ratio between the mean energy and the mean enstrophy, $\overline{E} / \overline{\Gamma}_2$, as a function of inverse temperature, $\beta$, in a square domain, $a=1$. The wavenumbers, $k$ and $l$, are truncated at $k_{max}$ and $l_{max}$, respectively. The total mode in the system is thus $N = k_{max} l_{max}$. As we increase the truncation wavenumber while fixing $\beta$, greater amounts of energy are distributed in higher wavenumbers, which results in the reduction of $\overline{E} / \overline{\Gamma}_2$. Conversely, when we decrease $\beta$ towards the critical value, $-\lambda_1 = - 2\pi^2 \fallingdotseq -19.74$, the energy concentrates in the lowest mode, $(k, l) = (1,1)$. Consequently, $\overline{E} / \overline{\Gamma}_2$ finally converges to $1 / \lambda_1 \fallingdotseq 0.0507$, regardless of the values of $k_{max}$ and $l_{max}$. Thus, a sharper peak of $\overline{E} / \overline{\Gamma}_2$ forms at $\beta = -\lambda_1$ for larger values of $(k_{max}, l_{max})$.}\label{fig:caloric}
\end{figure}

The model system adopted here is a two-dimensional flow contained in a rectangular domain with an area of unity, $0 \leq x \leq \sqrt{a}, \ 0 \leq y \leq 1/\sqrt{a}$, surrounded by rigid boundaries. The eigenvalues and eigenfunctions of the Laplacian operator are
\begin{align}
\lambda_{\boldsymbol{n}} & = \frac{\pi^2 k^2}{a} + \pi^2 a l^2 \label{eq:eigen_values} \\
e_{\boldsymbol{n}}(x, y) & = 2 \sin(\pi kx / \sqrt{a}) \sin(\pi \sqrt{a} ly) , \label{eq:eigenfunction}
\end{align}
where a vector index $\boldsymbol{n} = (k, l)$ with $k = 1, 2, \ldots$, and $l = 1, 2, \ldots$ is used. If required, the eigenvalues can be rearranged with a scalar index that yields $\{ \lambda_1, \lambda_2, \ldots \}$ in ascending order as (\ref{eq:lambda-order}). Regardless of the value of $a$ the smallest eigenvalue always corresponds to the $(k,l) = (1,1)$ mode; specifically, $\lambda_1 = \lambda_{11} = \pi^2(1 / a + a )$. After the wavenumbers, $k$ and $l$, are truncated at some numbers, $k_{max}$ and $l_{max}$, respectively, following the formulation in Section~\ref{sec:statistical}, the GCE with the total number of modes, $N = k_{max} l_{max}$, can be readily computed. In FIG.~\ref{fig:caloric}, we show $\overline{E} / \overline{\Gamma}_2$ as a function of $\beta$. The peak of $\overline{E} / \overline{\Gamma}_2$ at $\beta = - \lambda_1$ becomes sharper with increasing $N$. These plots demonstrate that all the possible states of any specified $\overline{E} / \overline{\Gamma}_2$ will reduce to a single value of $\beta = -\lambda_1$ in the thermodynamic limit. Thus, the finite truncation of the wavenumber is essential to allow for variations in $\beta$.

\begin{figure}
\centering
\includegraphics[height=8cm]{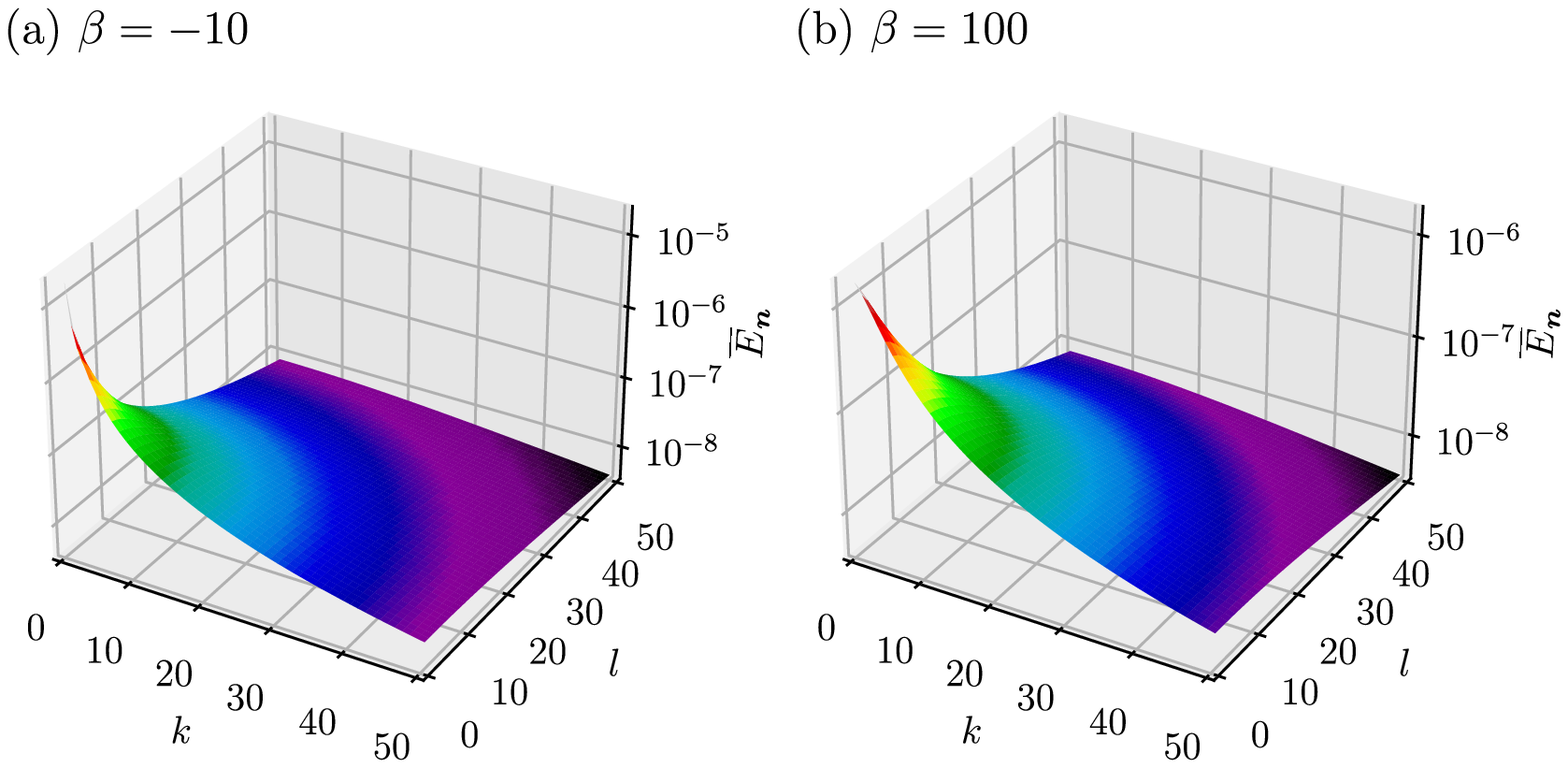}
\caption{Mean energy spectra of the grand-canonical ensemble (GCE). We have chosen a square domain, $a=1$, and truncate the wavenumber as $(k_{max}, l_{max}) = (50, 50)$. The inverse temperature is (a) $\beta = -10$ and (b) $\beta = 100$.}\label{fig:spectrum_gc}
\end{figure}

The expectation value of the energy spectrum, $\overline{E}_{\boldsymbol{n}}$, can now be represented as
\begin{equation*}
\overline{E}_{\boldsymbol{n}} = \frac{1}{2N(\beta + \lambda_{\boldsymbol{n}})} = \frac{1}{2N(\beta + \pi^2 k^2 / a + \pi^2 a l^2)} .
\end{equation*}
Function $\overline{E}_{\boldsymbol{n}}$ for a square domain, $a=1$, is plotted in FIG.~\ref{fig:spectrum_gc} for (a) negative ($\beta = -10$) and (b) positive ($\beta = 100$) temperatures. When $\beta$ is small and even close to the critical value, $-\lambda_1$, most energy concentrates in the lowest mode, $(k, l) = (1, 1)$. As we increase $\beta$, the energy becomes broadly distributed. In the asymptotic limit of $\beta \gg \lambda_N$, we obtain $\overline{E}_{\boldsymbol{n}} \sim 1 /(2N\beta)$; specifically, the energy is equally distributed in mode space.

\begin{figure}
\centering
\includegraphics[width=\textwidth]{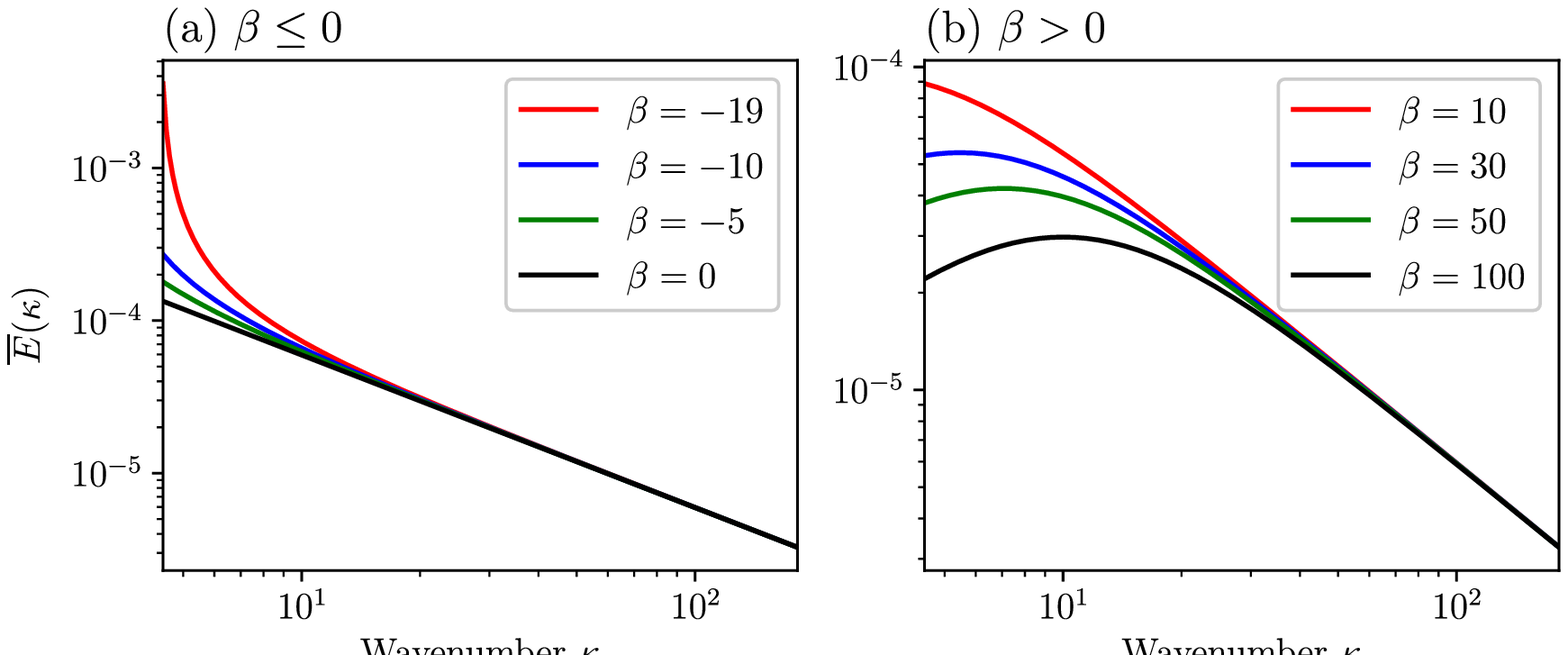}
\caption{One-dimensional energy spectra corresponding to the GCE, specified by (\ref{eq:1d-spectrum}), for (a)$\beta=-19, -10, -5, 0$ and (b)$\beta = 10, 30, 50, 100$. In these plots, the lower end of the $\kappa$-axis is chosen as $\sqrt{\lambda_1} \fallingdotseq 4.44$.}\label{fig:spectrum_1d_gce}
\end{figure}

Again assuming a square domain, $a = 1$, we define the absolute value of the wavenumber, $\kappa \equiv \sqrt{\lambda_{\boldsymbol{n}}} = \pi\sqrt{k^2 + l^2}$, as well as the azimuth angle, $\theta = \arctan(l / k)$; we regard them as continuous variables. Upon integration of the two-dimensional energy spectrum with respect to $\theta$, we derive the corresponding one-dimensional spectrum:
\begin{equation}
\overline{E}(\kappa) \propto \frac{\kappa}{\beta + \kappa^2} \ \ \text{with} \ \ \kappa \geq \sqrt{\lambda_1} .
\label{eq:1d-spectrum}
\end{equation}
Importantly, the treatment of $\kappa$ and $\theta$ as continuous variables is a crude approximation. The wavenumber discreteness cannot be ignored, particularly for a small-wavenumber component, $\kappa^2 \gtrsim \lambda_1$. Nevertheless, the simple expression (\ref{eq:1d-spectrum}) is useful for determining the dependence of the energy spectrum on the inverse temperature, $\beta$. As shown in FIG.~\ref{fig:spectrum_1d_gce}, when $- \lambda_1 < \beta < \lambda_1$, $\overline{E}(\kappa)$ is a monotonically decreasing function. In contrast, when $\beta > \lambda_1$, $\overline{E}(\kappa)$ has a maximum at $\kappa = \sqrt{\beta}$. Roughly, $\beta$ controls the energy concentration in the low-wavenumber components.

We now vary the aspect ratio $a$ over time. This situation corresponds to the case considered in Example I of Section~\ref{sec:formulation}. Here, the stream function for the boundary-induced potential flow, $\psi_p$, is specified as (\ref{eq:pure-strain}). The temporal variation of the Laplacian eigenfunction can be written as
\begin{align*}
\frac{\partial e_{\boldsymbol{n}}}{\partial t} = & - \pi k a^{-3/2} \dot{a} x \cos \left( \pi kx / \sqrt{a} \right) \sin \left( \pi \sqrt{a} ly \right) \nonumber \\
+ & \pi l a^{-1/2} \dot{a} y \sin \left( \pi kx / \sqrt{a} \right) \cos \left( \pi \sqrt{a} ly \right) \nonumber \\
= & \psi_{p,y} e_{\boldsymbol{n},x} - \psi_{p,x} e_{\boldsymbol{n},y} \nonumber .
\end{align*}
This expression shows that the coupling coefficients between the vortical flow and the potential flow, defined as $A_{nm}$ in (\ref{eq:coupling_A}), identically vanish. In other words, the boundary-induced flow does not cause any vorticity redistribution in mode space. The system is formally governed by the usual form of the vorticity equation:
\begin{equation*}
\dot{\omega}_{\boldsymbol{n}} = \sum_{\boldsymbol{m}, \boldsymbol{l}} B_{\boldsymbol{n}\boldsymbol{m}\boldsymbol{l}} \omega_{\boldsymbol{m}} \omega_{\boldsymbol{l}} ,
\end{equation*}
although the coupling coefficient $B_{\boldsymbol{n}\boldsymbol{m}\boldsymbol{l}}$ now changes over time.

In this system, energy contained in each mode varies according to
\begin{equation}
\dot{E_{\boldsymbol{n}}} = - \frac{\dot{\lambda}_{\boldsymbol{n}}}{\lambda_{\boldsymbol{n}}} E_{\boldsymbol{n}} + \frac{\omega_{\boldsymbol{n}}}{\lambda_{\boldsymbol{n}}} \sum_{\boldsymbol{m}, \boldsymbol{l}} B_{\boldsymbol{n}\boldsymbol{m}\boldsymbol{l}} \omega_{\boldsymbol{m}} \omega_{\boldsymbol{l}} ,
\label{eq:energy_rec_mode}
\end{equation}
where the tendency of the Laplacian eigenvalue is $\dot{\lambda}_{\boldsymbol{n}} = \pi^2 \dot{a} \left( l^2 - k^2 / a^2 \right)$. When summing (\ref{eq:energy_rec_mode}) for all $\boldsymbol{n}$, we obtain
\begin{equation}
\dot{E}_\omega = - \sum_{\boldsymbol{n}} \frac{\dot{\lambda}_{\boldsymbol{n}}}{\lambda_{\boldsymbol{n}}} E_{\boldsymbol{n}}.
\label{eq:energy_rec}
\end{equation}
Therefore, work performed by the boundary motion is attributed to the variations in the Laplacian eigenvalues, equivalently, the absolute values of the wavenumber. Work performed on the system is thus written as
$W = E_\omega(t) - E_\omega(0) = - \int^t_0 \sum_{\boldsymbol{n}}(\dot{\lambda}_{\boldsymbol{n}} / \lambda) E_{\boldsymbol{n}} dt'$.
The tendency of the enstrophy contained in each mode is similarly derived as
\begin{equation}
\frac{d}{dt} \left( \frac{\omega^2_{\boldsymbol{n}}}{2} \right) = \sum_{\boldsymbol{m}, \boldsymbol{l}} B_{\boldsymbol{n}\boldsymbol{m}\boldsymbol{l}} \omega_{\boldsymbol{n}} \omega_{\boldsymbol{m}} \omega_{\boldsymbol{l}} .
\label{eq:enstrophy_rec_mode}
\end{equation}
In contrast to the energy case, the total enstrophy is conserved: $\dot{\Gamma}_2 = 0$.

\subsection{Experiments}\label{experiments}

\subsubsection{Calculation conditions}\label{calculation-conditions}

We integrate the vorticity equation (\ref{eq:v-eq}) in a rectangular domain using eigenfunction expansion with basis functions of (\ref{eq:eigenfunction}). The nonlinear terms are evaluated in real space using a fast Fourier transform scheme. Aliasing errors are eliminated by truncation of the wavenumber based on the 3/2 rule. A total of $N = 41 \times 41 = 1681$ modes are included in the calculation. We adopt the third-order Runge-Kutta scheme developed by \citet{spalart_spectral_1991} for numerical integration. The aspect ratio of the domain shape is varied as $a(t) = a_0^{\sin(2 \pi t / \tau)}$ with $a_0 = 4$, which induces a pure straining potential flow described as (\ref{eq:pure-strain}). Accordingly, the domain shape is restored to its initial state, $a = 1$, at $t = \tau / 2, \tau, 3\tau / 2, \ldots$. The initial conditions of the vorticity field, $\left\{ \omega_{\boldsymbol{n}} \right\}$, are sampled from a GCE. Because each mode follows a Gaussian distribution and is statistically independent, non-biased samples can easily be arranged from uniform random numbers using the Box-Muller method. Here, we designate the number of samples as $M$. In this study, we vary $M$, the inverse temperature $\beta$, and the distortion time scale $\tau$ to see the dependence of the result on each parameter. The results of varying $M$ are provided in the Appendix \ref{appendix B}. Although thousands of samples are desirable to obtain a robust statistical estimate, because of limited computational resources, we adopt $M=8000$ for only the first case and subsequently employ $M=576$ to discuss the \yo{dependence} on $\beta$ and $\tau$. Inevitably, non-negligible random errors may be contained in each experimental result.

In addition to $\tau$, the system involves a typical time scale $L / U$, where $L$ and $U$ represent the domain size and the typical velocity of the vortical flow part, respectively. This time scale roughly corresponds to the period that the largest eddy requires to overturn. In the present model, because the domain area is set to $1$, $L$ may also be regarded as unity. We then define $\tau_e \equiv 1 / \sqrt{E_0}$, where $E_0$ is the mean initial energy determined by $\beta$ and $N$ as (\ref{eq:energy-exp}); we set the distortion time scale $\tau$ with reference to the eddy turnover time scale, $\tau_e$.

\subsubsection{Results}\label{calculation-conditions}

\begin{figure}
\centering
\includegraphics[height=8cm]{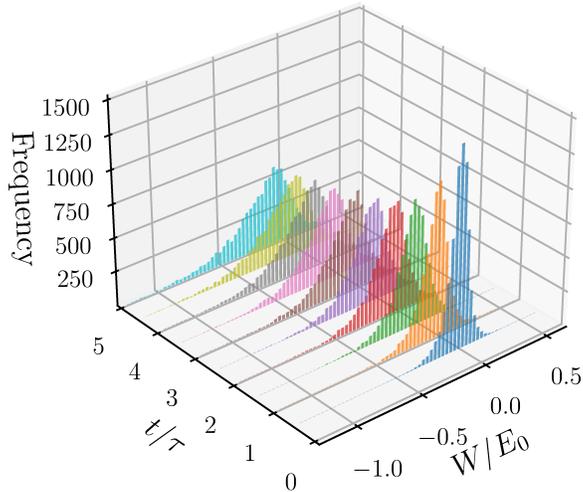}
\caption{Histograms of the cumulative work $W$ performed by the moving boundary on the system at the instances when the aspect ratio is unity (i.e., $t$ is a multiple of $0.5\tau$). The experimental parameters are $\beta=-18$, $\tau=\tau_e$ and $M=8000$.}
\label{fig:work_histogram}
\end{figure}

First, we focus on a particular case of $\beta = -18$, $M=8000$, and $\tau = \tau_e$. Pressure work performed by the boundary on the system is shown in FIG.~\ref{fig:work_histogram} as a histogram. In the initial stage of the experiment, the shape of the histogram is almost symmetric, and we cannot find a significant tendency for the sign of $W$. Over time, the histogram becomes negatively skewed; even the mode location shifts towards $W < 0$. Until $t = 5\tau$, $79.6\%$ of the samples received negative work from the boundary. The supplemental material shows a movie of colored contours of the stream function, $\psi_\omega$, for a specific sample.

\begin{figure}
\centering
\includegraphics[width=0.8\textwidth]{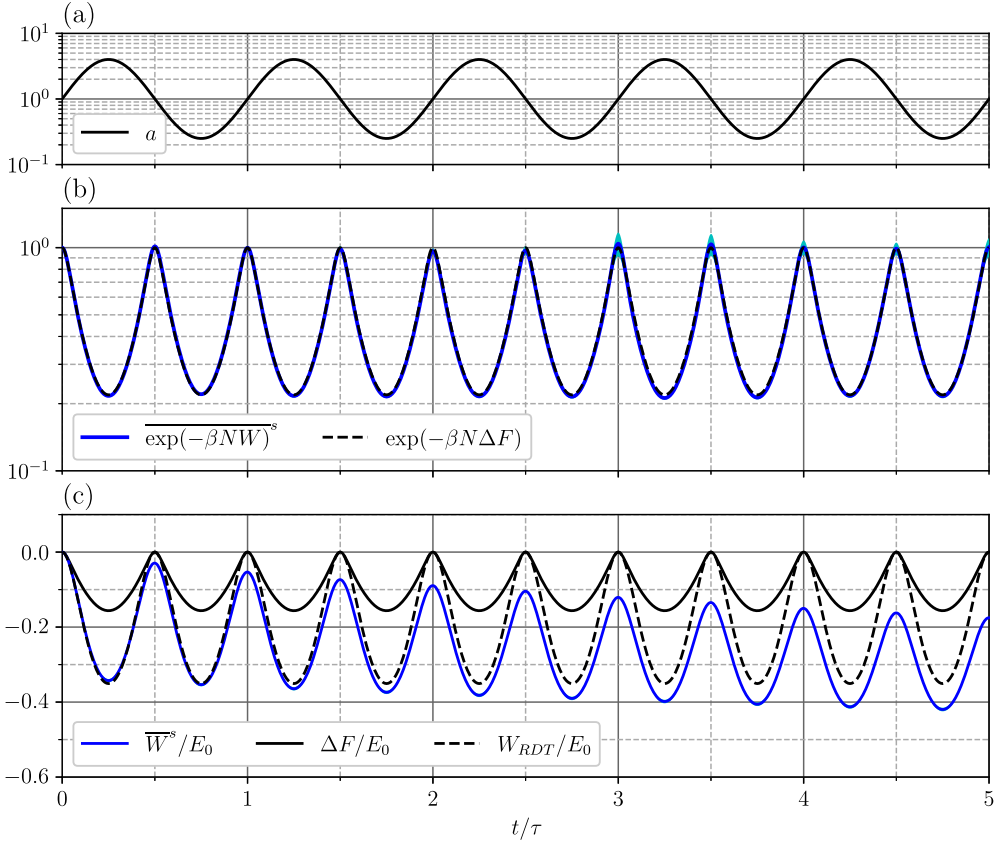}
\caption{Results of the experiment for $\beta = -18$, compared with theoretical estimates. (a) Time series of the aspect ratio, $a$. (b) The blue curve represents $\overline{\exp(-N \beta W)}^s$, with the standard error indicated by cyan. The dashed curve is the analytically derived $\exp(-N \beta \Delta F)$. (c) The blue curve represents the mean work performed on the system, $\overline{W}^s$. The standard error is indicated by cyan, although it is almost invisible. The black solid curve is $\Delta F$ and the black dashed curve is the work predicted by the rapid distortion theory.}\label{fig:time_series_work}
\end{figure}

In the following text, we use an overline with a superscript $s$ to represent a sample mean of a variable. FIG.~\ref{fig:time_series_work}b compares the experimental results, $\overline{\exp(-N \beta W)}^s$, with the theoretical prediction, $\exp(-N \beta \Delta F)$, where $\Delta F$ is the difference in the free energy given by
\begin{equation*}
\Delta F = - \left. \frac{1}{2N\beta} \sum_{\boldsymbol{n}} \ln \left( \frac{\lambda_{\boldsymbol{n}}(t')}{\beta + \lambda_{\boldsymbol{n}}(t')} \right) \right|_0^t .
\end{equation*}
The theoretical estimates and the experimental results are in good agreement. The Jarzynski equation (\ref{eq:Jarzynski}) for this system is thus validated. FIG.~\ref{fig:time_series_work}c shows the time series of the sample mean of work performed on the system, $\overline{W}^s$. The difference in the free energy, $\Delta F$, and the work predicted by RDT are also shown. Here, the prediction from RDT corresponds to the results obtained if the cross-mode coupling is neglected. In the initial stage, $t / \tau < 1$, RDT adequately explains the periodic variation in the mean energy. However, $\overline{W}^s$ gradually decreases in later periods and never returns to 0. We also confirm the relationship between the mean work performed on the system and the difference in the free energy, $\overline{W}^s < \Delta F$, predicted from the Jarzynski equality under the condition of $\beta < 0$.

\begin{figure}
\centering
\includegraphics[width=\textwidth]{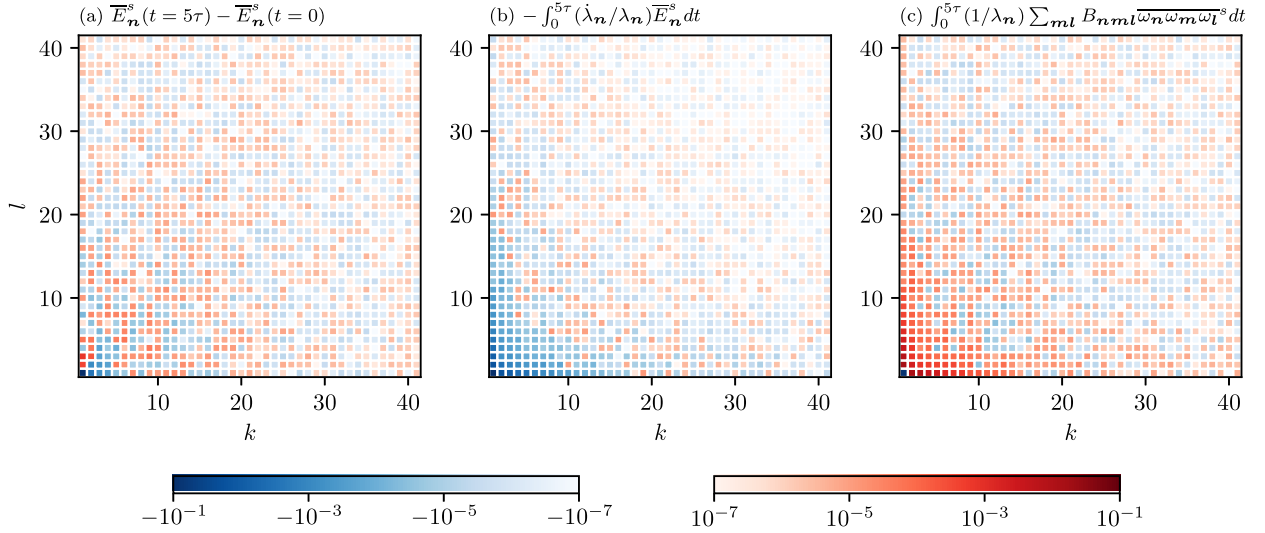}
\caption{Two-dimensional spectral plots of the energy budget for a $M=8000$, $\beta=-18$ and $\tau = \tau_e$ experiment. (a) Net variations in the mean energy spectrum, $\overline{E}^s_{\boldsymbol{n}}(t = 5\tau) - \overline{E}^s_{\boldsymbol{n}}(t = 0)$. (b) Temporally integrated energy production term, $- \int_0^{5\tau} (\dot{\lambda}_{\boldsymbol{n}}/\lambda_{\boldsymbol{n}}) \overline{E}^s_{\boldsymbol{n}} dt$. (c) Temporally integrated nonlinear interaction term, $\int_0^{5\tau} (1/\lambda_{\boldsymbol{n}}) \sum_{\boldsymbol{m}\boldsymbol{l}} B_{\boldsymbol{n}\boldsymbol{m}\boldsymbol{l}} \overline{\omega_{\boldsymbol{n}} \omega_{\boldsymbol{m}} \omega_{\boldsymbol{l}}}^s dt$. All data are scaled by $E_0$, and the color is represented in a logarithmic scale for positive and negative values, respectively.}\label{fig:energy_budget_2D}
\end{figure}

Next, we analyze the detailed energy budget in spectral space. Equation (\ref{eq:energy_rec_mode}) is averaged over the samples and integrated over $0 \leq t \leq 5\tau$; its terms for each $\boldsymbol{n} = (k, l)$ are compared in FIG.~\ref{fig:energy_budget_2D}. FIG.~\ref{fig:energy_budget_2D}b and c show that, in most of the low-wavenumber regions (specifically, $k + l \leq 10$), the energy production term is negative, while the nonlinear interaction term is positive. An exception is the lowest mode $\boldsymbol{n} = (1,1)$, in which both the production and nonlinear interaction terms are negative. These results indicate that energy is transferred from the lowest mode towards the adjacent higher modes. For a higher-wavenumber region, $k + l \gg 10$, the production and nonlinear interaction terms show random variations. The net variations of the energy spectrum in FIG.~\ref{fig:energy_budget_2D}a, which is the sum of the following two panels, exhibit significant energy loss in the lowest mode but no clear tendency to gain or loss in the other modes.

\begin{figure}
\centering
\includegraphics[width=0.8\textwidth]{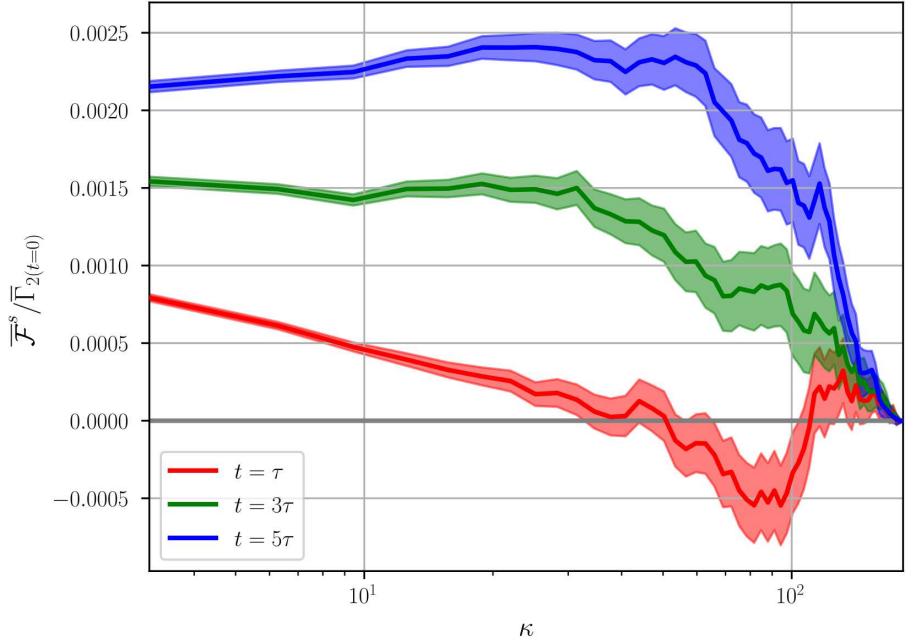}
\caption{Sample means of the enstrophy flux $\mathcal{F}(\kappa)$ in one-dimensional wavenumber space defined as (\ref{eq:enstrophy_flux}) at $t = \tau$, $3\tau$ and $5\tau$. The vertical axis is scaled by the expectation value of the total enstrophy at the initial time. Means and their standard errors are shown.}
\label{fig:enstrophy_flux}
\end{figure}

We also discuss the enstrophy budget in spectral space. In contrast to energy, enstrophy is not produced by the external force; only the nonlinear interaction term must be considered. Here, we define the cumulative enstrophy flux in one-dimensional wavenumber space as
\begin{equation}
\mathcal{F}(\kappa) = \sum_{|\boldsymbol{n}| > \kappa / \pi} \int_0^t \sum_{\boldsymbol{m} \boldsymbol{l}} B_{\boldsymbol{n}\boldsymbol{m}\boldsymbol{l}} \omega_{\boldsymbol{n}} \omega_{\boldsymbol{m}} \omega_{\boldsymbol{l}} dt' ,
\label{eq:enstrophy_flux}
\end{equation}
where $|\boldsymbol{n}| = \sqrt{k^2 + l^2}$. This flux corresponds to the gain of enstrophy for a high-mode region with a wavenumber greater than $\kappa$ until time $t$. Because the wavenumber of each mode varies as $\sqrt{\lambda_{\boldsymbol{n}}(t)}$, $\mathcal{F}$ does not strictly coincide with the usual definition of enstrophy flux in homogeneous turbulence. FIG.~\ref{fig:enstrophy_flux} shows $\mathcal{F}$ for $t = \tau, 3\tau$, and $5\tau$. We confirm that enstrophy contained in the lowest-mode component is transferred to higher wavenumbers. At earlier times, the enstrophy flux is constrained in the low-wavenumber region; a negative flux appears in the intermediate wavenumber range. In later periods, the positive flux approaches higher wavenumbers, and the negative flux is no longer confirmed.

The downscale enstrophy transfer is a natural consequence of the energy loss of the system. To explain this, we focus on the ratio between energy and enstrophy,
\begin{equation*}
\frac{\overline{E}_\omega}{\overline{\Gamma}_2} = \frac{\sum_{\boldsymbol{n}} \overline{\omega^2_{\boldsymbol{n}}} / \lambda_{\boldsymbol{n}}}{\sum_{\boldsymbol{n}} \overline{\omega^2_{\boldsymbol{n}}}},
\end{equation*}
formally equivalent to the centroid of $1/\lambda_{\boldsymbol{n}}$ weighted by the enstrophy. Because $\overline{\Gamma}_2$ is conserved, the gain or loss of energy is accompanied by an increase or decrease in this centroid, corresponding to upscale or downscale enstrophy transfer, respectively. Two-dimensional turbulence is generally associated with the upscale and downscale cascades of energy and enstrophy, respectively \citep{fjortoft_changes_1953, kraichnan_inertial_1967}. Loss of energy via the pressure force at the lateral boundary plays a similar role to the upscale energy transfer in the downscale enstrophy cascade.

\begin{figure}
\centering
\includegraphics[width=0.8\textwidth]{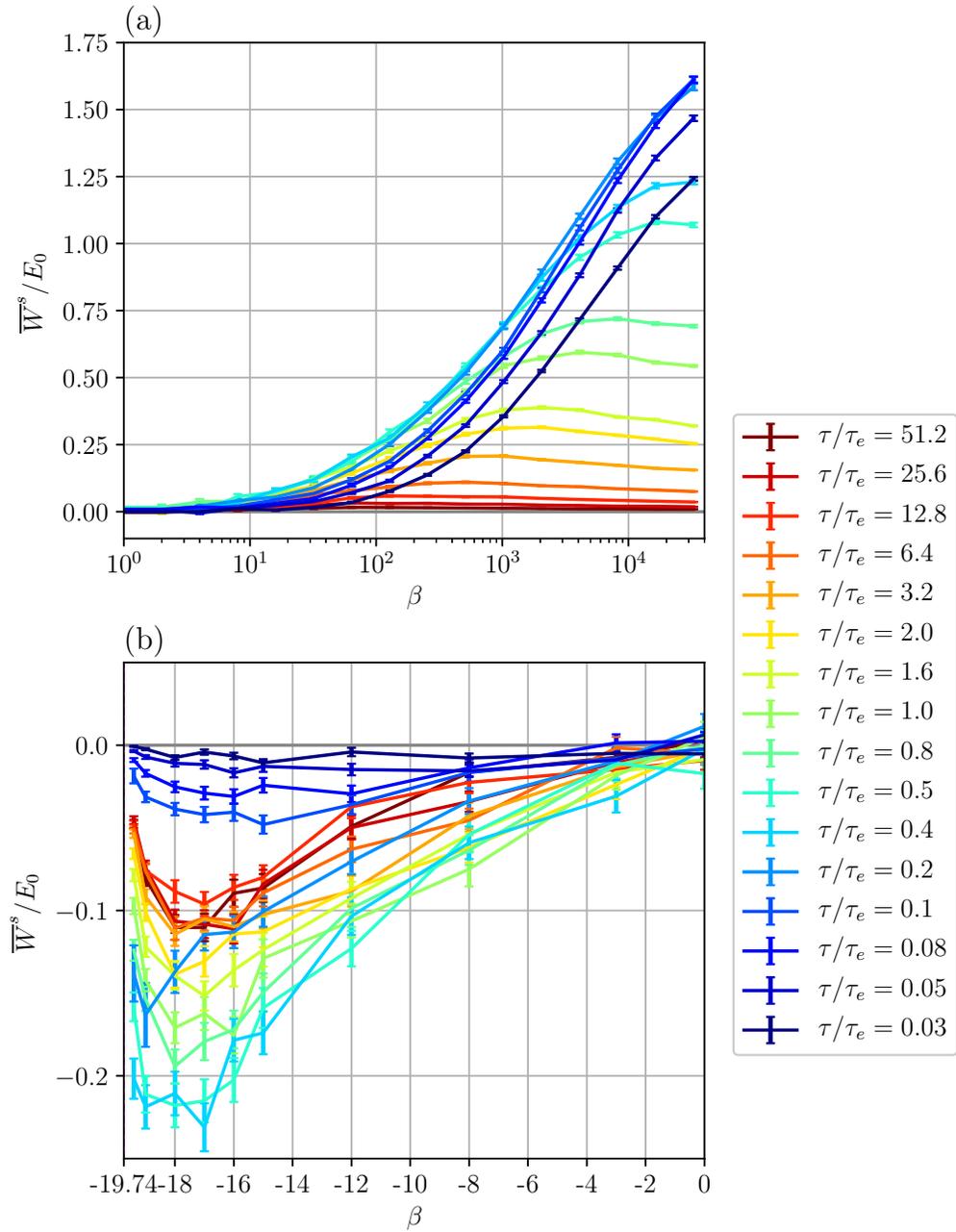}
\caption{Sample means of the work performed on the system scaled by the initial mean energy, $\overline{W}^s / E_0$, at $t = 5\tau$ against $\beta$ for (a) $\beta > 0$ and (b) $\beta \leq 0$. The horizontal axis in (a) uses a logarithmic scale while the horizontal axis in (b) uses a linear scale. Error bars represent the standard errors.}
\label{fig:varying_beta}
\end{figure}

The Jarzynski equality indicates that the direction of the energy transfer through the lateral boundary depends on the sign of $\beta$. We validate this prediction by carrying out experiments for various $\beta$ ranges from negative to positive values. The obtained $\overline{W}^s / E_0$ are shown in FIG.~\ref{fig:varying_beta}. In accordance with the theoretical expectation, the sign of $\overline{W}^s$ switches with the sign of $\beta$. When we decrease $\beta$ until $-\lambda_1$, all the energy is concentrated at the lowest mode, $(k, l) = (1, 1)$. Such a single eddy at the largest scale is a stable solution of the Euler equation. Consequently, no enstrophy is transferred towards higher modes. The flow field evolves according to the RDT; the distortion process is completely reversible. Therefore, $\overline{W}$ should converge to $0$ in this limit. In addition, with the assumption that $\overline{W}$ continuously changes the sign between positive and negative temperature regions, $\overline{W} = 0$ should hold at $\beta = 0$. Consequently, we predict that $\overline{W} / E_0$ will have a minimum value between $\beta = -19.74$ and $\beta = 0$. FIG.~\ref{fig:varying_beta} shows that this minimum is located in $- 19 \leq \beta \leq -15$ for every $\tau/\tau_e$ case.

\clearpage

Next, we focus on the \yo{dependence} of $\overline{W} / E_0$ on the distortion time $\tau$. First, in the asymptotic limit of $\tau / \tau_e \ll 1$, the system's behavior may be adequately predicted by the RDT. Because the RDT describes a completely reversible process, it follows that $\overline{W} / E_0 = 0$ for $\tau \to 0$. The opposite side of the asymptotic limit is a very slow distortion, $\tau / \tau_e \gg 1$. For conventional thermodynamic theory, this situation corresponds to the quasistatic process, in which the probability density, $P(\boldsymbol{\omega})$, is always very close to an equilibrium state; consequently, the process is reversible. In the present model, however, the existence of such an equilibrium state at each time is uncertain. We thus cannot conclude that $\overline{W} / E_0 = 0$ for $\tau \to \infty$ based on theoretical reasoning.

\begin{figure}
\centering
\includegraphics[width=0.8\textwidth]{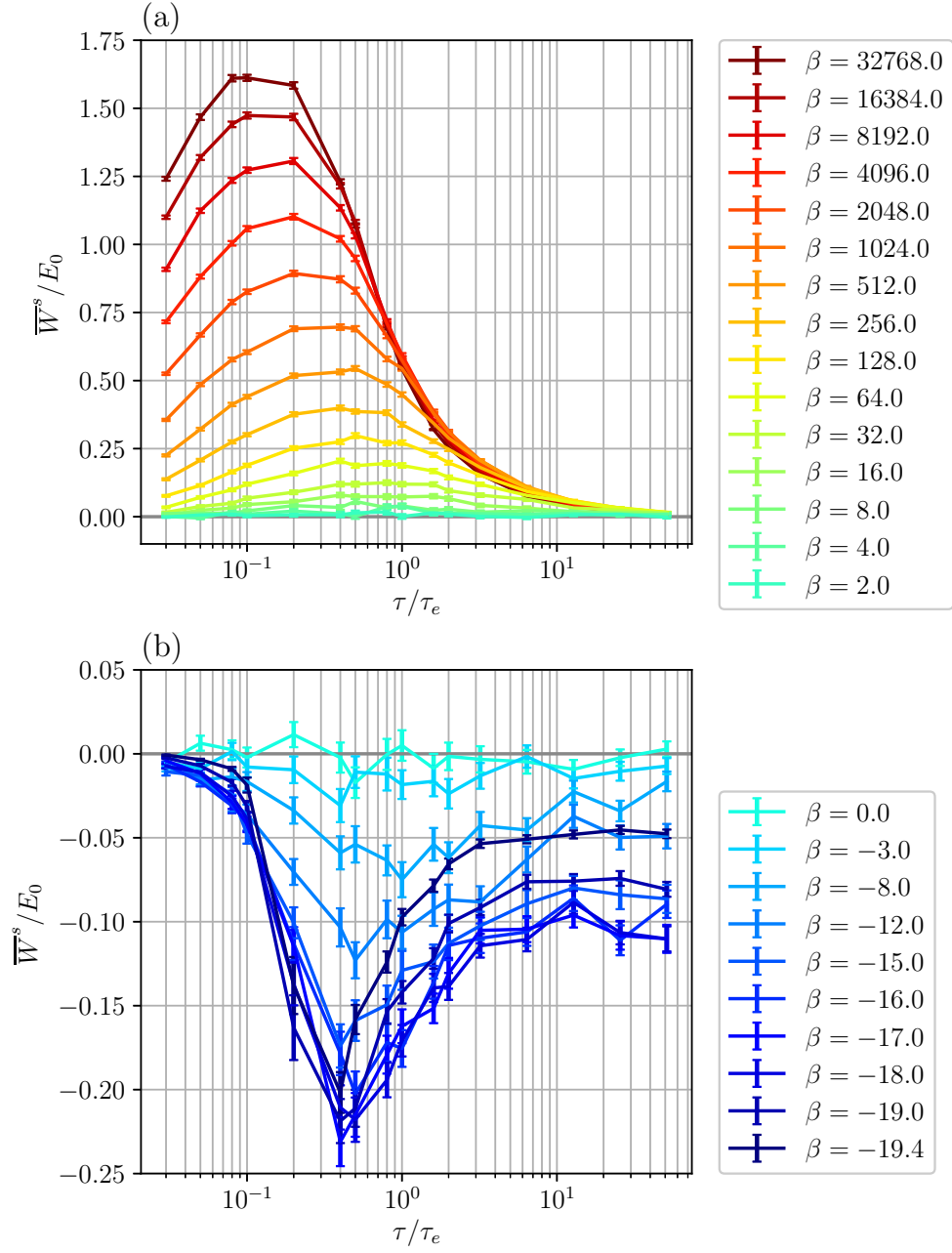}
\caption{Sample means of the work performed on the system scaled by the initial mean energy, $\overline{W}^s / E_0$, at $t = 5\tau$ against $\tau / \tau_e$ for (a) $\beta > 0$ and (b) $\beta \leq 0$. The error bars represent the standard errors.}\label{fig:varying_tau}
\end{figure}

The plots of $\overline{W}^s / E_0$ against $\tau$ are shown in FIG.~\ref{fig:varying_tau}. Indeed, $\left| \overline{W}^s \right| / E_0$ decreases monotonically towards $0$ as we decrease $\tau / \tau_e$ for $\tau / \tau_e < 0.1$. In contrast, the results separate for a large $\tau / \tau_e$, depending on the sign of $\beta$. Although $|\overline{W}^s| / E_0$ decreases monotonically towards $0$ for positive $\beta$, it does not decrease in such a manner for negative $\beta$. Apparently, $\overline{W}^s / E_0$ approaches a constant value prescribed by $\beta$, but a proper explanation of this unexpected result has not yet been provided. Additionally, for negative $\beta$, the most efficient energy extraction occurs at $0.4 \leq \tau / \tau_e \leq 1$. In contrast, for positive $\beta$, the location of the peak in $\overline{W}^s / E_0$ shifts towards lower $\tau / \tau_e$ as $\beta$ increases. This result can be rationalized by focusing on the turbulence length scale. As shown in FIG.~\ref{fig:spectrum_1d_gce}, when we increase $\beta$, a larger part of the energy is distributed to higher wavenumbers. Accordingly, the typical length scale of the turbulent eddies decreases; this leads to a considerably shorter turnover time of eddies compared with the original estimate, $\tau_e = 1 / \sqrt{E_0}$. The distortion time and turnover time should be comparable for the potential flow induced by domain distortion to interfere most efficiently with turbulent eddies. Therefore, for large $\beta$, efficient energy injection occurs at smaller $\tau / \tau_e$.

\clearpage

\section{Discussion and conclusions}\label{sec:conclusions}

In most physical systems, the second law of thermodynamics restricts the amount of work available from microscopic random motions. More specifically, heat energy cannot be extracted from an equilibrium state without changes to any other macroscopic condition. However, this principle is valid only when the absolute temperature is positive. In a negative temperature state, macroscopic operations irreversibly extract heat energy from the system. A two-dimensional flow is a unique system in which the temperature can be negative. \yo{The simple model used here\textemdash incompressible Euler equation system surrounded by a moving lateral boundary\textemdash allows us to theoretically and numerically explore the energetics of negative temperature states of turbulent flows. A specific choice of initial conditions of statistical equilibria employed from Kraichnan's classical work enables direct application of Jarzynski's formula. We have thus derived, as expected, a simple expression that exhibits the one-way energy transfer from turbulent fluctuations to macroscopic pressure work.}

% This study introduces a novel conceptual model, incompressible Euler equation system surrounded by a moving lateral boundary, which allows us to theoretically and numerically explore the energetics of negative temperature state of turbulent flows. A specific choice of initial conditions of statistical equilibria employed from Kraichnan's classical work enables direct application of Jarzynski's formula. We have thus derived, as expected, a simple expression that exhibits the one-way energy transfer from turbulent fluctuations to macroscopic pressure work.

Several caveats exist, however, as to the initial equilibrium state adopted in this study. As is the common notion in statistical mechanics, a thermal equilibrium would be regarded as a state that a system reaches after a sufficiently long time has passed from an arbitrary state. The present Kraichnan's GCE is not genuine in that sense; frankly, it is a particular class of invariant measure of the truncated Euler equation. An assumption on the ergodicity of the truncated Euler equation would provide a micro-canonical ensemble formulated by Bouchet and Corvellec \cite{bouchet_invariant_2010} as a better model, which still ignores constraints from higher-order Casimir invariants.
To formulate a model much more strictly, taking the continuous limit by setting the truncation wavenumber as infinite and considering all the Casimir invariants to define the equilibria \yo{are} desired. This procedure is nothing but the MRS theory. Such a rigorous method, however, provides only a trivial result as to the response of the system to the external boundary forcing. It is because the MRS theory predicts the macroscopic flow state in a completely deterministic form. Accordingly, the work's probability distribution reduces to a delta function on the mean state. A Jarzynski-like formulation will not be applicable in this line. Intriguing results essentially originate from turbulent fluctuations that are allowed to reside in an equilibrium for a system with a large but finite number of elements. These explain why the standard approach to defining equilibria for a continuous flow system was abandoned in the present study.

Although the validity of the initial conditions of equilibrium requires further investigation, interpreting the results of numerical analysis is rather straightforward. If the temperature is negative, a large amount of energy and enstrophy are concentrated at the lowest-wavenumber mode. When one disturbs the system by moving the domain boundary, enstrophy in the lowest mode is redistributed towards higher wavenumbers. This enstrophy transfer is inevitably accompanied by the energy loss of the system. We have seen that the energy and enstrophy budgets in spectral space obtained from an experiment support this mechanism. If the temperature is increased to positive values, since the enstrophy concentration to the lowest mode is no longer prominent, an inverse enstrophy cascade becomes possible; thus, the system can gain energy. A series of parameter-sweep experiments further show that the energy exchange rate is dependent on temporal parameters. Specifically, the most efficient energy exchange occurs when the two time scales, those of cyclic boundary motion and the overturning of turbulent eddies, are comparable. These intuitively reasonable results are likely not an artifact originating from Kraichnan's GCE but a ubiquitous feature of statistical equilibria of two-dimensional Euler equation systems.

We shall discuss the present results in connection with the existing knowledge of two-dimensional flows. Commonly in various Earth and planetary systems, coherent flow structures such as jets or vortices spontaneously emerge from a turbulent state. Energy conversion from microscopic fluctuations to a macroscopic structure plays a driving role in this process. From this perspective, it is possible to regard the extraction of turbulence energy to an externally imposed potential flow as a derivative of the structure-formation mechanisms.

This paper has considered a simple Euler equation, but a similar approach applies to quasi-geostrophic systems that involve planetary-beta effects and variations in layer thicknesses. Furthermore, we can even replace the rigid lateral boundary by periodic conditions. In this case, variation in the domain shape is represented by decomposing a velocity vector into potential and vortical flow parts, $\boldsymbol{u} = \boldsymbol{u}_\omega + \boldsymbol{u}_p$, and imposing the stream function of the potential flow part. For example, $\psi_p$ defined as Eq. (\ref{eq:pure-strain}) induces the variation in the aspect ratio of the domain. By combining different boundary configurations with a more general quasi-geostrophic equation, we can discuss energy exchanges among jets, vortices, Rossby waves, and small-scale turbulence interacting with an externally imposed potential flow. For any kind of situations, the sign of temperature, if defined, is presumably a factor that determines the energy gain or loss of the system.

To make a model relevant to the real geophysical and astrophysical situations, it is important above all to formulate a statistical theory meaningful even in the thermodynamic, or equivalently continuous, limit. For this purpose, a plausible way is to include the diabatic source and sink of vorticity and net energy transfer between forcing and dissipation scales. Forced-dissipated turbulence of two-dimensional flows has also been explored in many other works (see \citep{bouchet_statistical_2012} and references therein). Recent studies have focused on the mechanisms of emergence and maintenance of a coherent flow structure much greater than forcing scales (e.g., \citep{srinivasan_zonostrophic_2012, marston_generalized_2016, frishman_jets_2017, woillez_theoretical_2017, frishman_turbulence_2018}). Future research should extend discussion to a generic non-equilibrium state of turbulence.

Finally, although this study has considered statistical quantities obtained by averaging over innumerable samples, it is also important to concentrate on a specific realization and to follow its transitions. In such problems, we occasionally find a drastic change in the flow structure over a short period, which bridges bistable states \citep{bouchet_langevin_2014, bouchet_random_2009, bouchet_control_2011, laurie_computation_2015, bouchet_rare_2019, herbert_atmospheric_2020}. Temporal variation in the domain geometry is expected to switch a stable state to an unstable state, thus triggering an abrupt transition of the flow structure to another state, like the vapor-liquid transition induced by compression or expansion of a fluid volume. Indeed, \citet{yasuda_new_2017} suggested that the splitting of a polar vortex in the stratosphere can be explained in terms of a phase transition induced by variation in a geometrical condition. This paper will hopefully motivate discussions of the statistical mechanics of geophysical fluid in those directions.

\section*{Acknowledgments}\label{acknowledgments}

The author thanks Sylvain Joubaud and Corentin Herbert for carefully reading the manuscript and suggesting informative, relevant literature. The author is also grateful to three anonymous reviewers for giving critical comments that significantly improved the manuscript. This research was supported by JSPS Overseas Research Fellowship and KAKENHI Grants JP18H04918 and JP20K14556. Computations were carried out using the Fujitsu PRIMERGY CX600M1/CX1640M1 (Oakforest-PACS) at the Information Technology Center of the University of Tokyo. Computations were also performed using the computer resources offered under the category of General Projects by Research Institute for Information Technology, Kyushu University.

\appendix

\section{Temporal differentiation of a spatially integrated quantity}\label{appendix A}

In the present model, since the domain geometry varies with time, special care is required to compute the temporal differentiation of an integrated quantity. To see this, let us first introduce a real continuous function $C(\boldsymbol{r}, t)$ that is positive inside of $\mathcal{D}$ and negative outside of $\mathcal{D}$. Clearly, the boundary of the domain $\partial \mathcal{D}$ is specified by the condition, $C = 0$. Accordingly,
\begin{align} \label{eq:boundary_C}
\frac{\partial C}{\partial t} + \boldsymbol{u} \cdot \nabla C = 0
\end{align}
is always satisfied on $\partial \mathcal{D}$. Letting $s \equiv \lvert \nabla C \rvert^{-1} \partial C / \partial t$, (\ref{eq:boundary_C}) represents the kinematic boundary condition (\ref{eq:boundary}).

Now, using the Heaviside function $H$, the integration of an arbitrary function $f(\boldsymbol{r}, t)$ over $\mathcal{D}$ is expressed as
\begin{align*}
\int_{\mathcal{D}} f d\boldsymbol{r} = \int_{\mathbb{R}^2} H(C(\boldsymbol{r}, t)) f(\boldsymbol{r}, t) d\boldsymbol{r} .
\end{align*}
The temporal differentiation of this expression is
\begin{align} \label{eq:diff_f}
\frac{d}{dt} \int_{\mathcal{D}} f d\boldsymbol{r} = \int_{\mathbb{R}^2} \frac{\partial C}{\partial t} \delta (C(\boldsymbol{r}, t)) f d\boldsymbol{r} + \int_{\mathbb{R}^2} H(C(\boldsymbol{r}, t)) \frac{\partial f}{\partial t} d\boldsymbol{r} ,
\end{align}
where $\delta$ is the Dirac's delta function. Since the integration of the first term on the right-hand side is contributed only from $\partial \mathcal{D}$ where $C = 0$, we may use (\ref{eq:boundary_C}) to rewrite it as
\begin{align*}
\int_{\mathbb{R}^2} \frac{\partial C}{\partial t} \delta (C(\boldsymbol{r}, t)) f d\boldsymbol{r} & = - \int_{\mathbb{R}^2} \boldsymbol{u} \cdot \nabla C \delta (C(\boldsymbol{r}, t)) f d\boldsymbol{r} \\
& = - \int_{\mathbb{R}^2} \boldsymbol{u} \cdot \nabla H(C(\boldsymbol{r}, t)) f d \boldsymbol{r} \\
& = \int_{\mathbb{R}^2} H(C(\boldsymbol{r}, t)) \boldsymbol{u} \cdot \nabla f d \boldsymbol{r}
\end{align*}
where we have used the incompressible condition, $\nabla \cdot \boldsymbol{u} = 0$, and the integration by parts. Inserting this to (\ref{eq:diff_f}), we finally derive
\begin{align} \label{eq:integrate_f}
\frac{d}{dt} \int_{\mathcal{D}} f d\boldsymbol{r} & = \int_{\mathbb{R}^2} H(C(\boldsymbol{r}, t)) \left[ \frac{\partial f}{\partial t} + \boldsymbol{u} \cdot \nabla f \right] d \boldsymbol{r} \nonumber \\
& = \int_{\mathcal{D}} \left[ \frac{\partial f}{\partial t} + \boldsymbol{u} \cdot \nabla f \right] d\boldsymbol{r} .
\end{align}
Expression (\ref{eq:integrate_f}) is useful to formulate the variations in energy or prove the conservation of Casimir invariants.

\section{Evaluation of errors}\label{appendix B}

For the GCE of a truncated Euler equation system, the deviation of a sampled macroscopic quantity such as $E_\omega$, $W$, or $\mathcal{F}$ from a genuine ensemble mean is massive. A robust estimate of a mean value requires a sufficiently large number of samples. Here, we consider the amount of work $W$ as an example. The confidence interval of the sample mean of $W$ obtained in experiments is represented as $(\overline{W}^s - c \delta W, \overline{W}^s + c \delta W)$, where $\overline{W}^s$ is the sample mean, $\delta W$ is the standard error of the sample mean, and $c$ is a constant. The standard error is defined as $\delta W = S / \sqrt{M}$, where $S$ is the sample standard deviation of $W$. For a normal distribution with $M \gg 1$, the 95\% confidence interval corresponds to $c = 1.96$, but for simplicity we use $c=1$.

\begin{figure}
\centering
\includegraphics[width=0.6\textwidth]{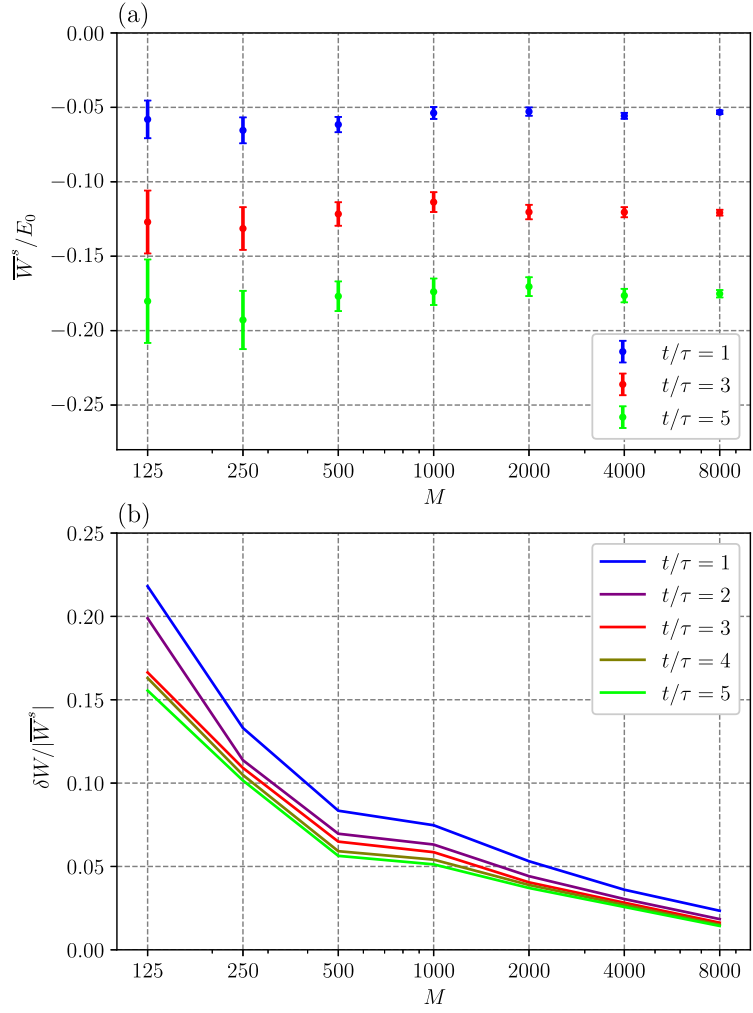}
\caption{(a)Estimated mean work performed on the system until $t = \tau, 3\tau, 5\tau$ for $\beta=-18$. The horizontal axis, $M$, is the number of samples. Error bar represents a standard error. (b)Sample standard errors of $W$ divided by the absolute values of sample means against $M$ for $\beta = -18$ and various $t$.}
\label{fig:varying_sample_number}
\end{figure}

To assess the \yo{dependence} of the results on the number of samples, we refer to FIG.~\ref{fig:varying_sample_number} for the estimated sample means and the standard errors of $W$ for various $M$ and $t$. We notice that the relative error of $W$ depends on the number of samples and time. We can estimate the rate of loss of energy from the system with more confidence in later periods. Overall, as is inferred from FIG.~\ref{fig:varying_sample_number}b, to suppress the relative error of the estimates of $W$ within 10\%, it would be sufficient to set $M$ greater than 500 for this case.

\clearpage

\bibliography{PR2022_onuki_revision}% Produces the bibliography via BibTeX.

\end{document}